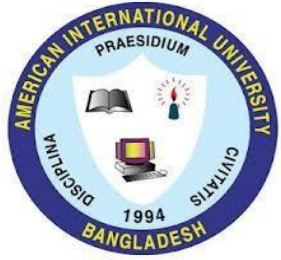

# Tissue Artifact Segmentation and Severity Analysis for Automated Diagnosis Using Whole Slide Images

by

**Galib Muhammad Shahriar Himel**

**(21-92094-2)**

Master of Science in Computer Science

(Intelligent Systems)

Department Computer Science

Faculty of Science and Technology

American International University-Bangladesh

September 2022

# Declaration

I declare that this thesis is my original work and has not been submitted in any form for another degree or diploma at any university or other institute of tertiary education. Information derived from the published and unpublished work of others has been acknowledged in the text and a list of references is given.

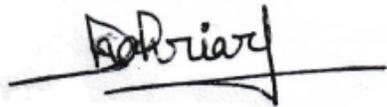

\-\-\-\-\-\-\-\-\-\-\-\-\-\-\-\-\-\-\-\-\-\-\-\-\-\-\-\-\-\-\-\-\-\-\-\-\-\-\-\-\-\-\-\-\-\-\-\-\-\-\-\-\-\-\-\-\-\-\-\-\-\-\-\-

**Galib Muhammad Shahriar Himel**

**21-92094-2**

# Approval

The thesis titled "Tissue Artifact Segmentation and Severity Analysis for Automated Diagnosis Using Whole Slide Images" has been submitted to the following respected members of the board of examiners of the department of computer science in partial fulfilment of the requirements for the degree of Master of Science in Computer Science on 15.09.2022 and has been accepted as satisfactory.

---------------------------------------------------

**Dr. Hossain Md. Shakhawat**

Assistant Professor & Supervisor

Department of Computer Science

American International University-Bangladesh

---------------------------------------------------

**Dr. Md. Sakir Hossain**

Associate Professor & External

Department of Computer Science

American International University-Bangladesh

---------------------------------------------------

**Dr. Tabin Hasan**

Professor & Head (Graduate Program)

Department of Computer Science

American International University-Bangladesh

---------------------------------------------------

**Dr. Dip Nandi**

Professor & Director (Computer Science)

Department of Computer Science

American International University-Bangladesh

---------------------------------------------------

**Dr. Tafazzal Hossain**

Professor & Dean (Computer Science)

Department of Computer Science

American International University-Bangladesh

---------------------------------------------------

**Dr. Carmen Z. Lamagna**

Vice Chancellor

American International University-Bangladesh

# Acknowledgement

First of all, I would like to express my gratitude to my thesis supervisor Dr. Hossain Md. Shakhawat, Assistant Professor, Department of Computer Science, American International University-Bangladesh (AIUB) for his adequate guidance and motivation in all aspects of my thesis work.

I would like to express my gratitude to my external supervisor, Dr. Md. Sakir Hossain, Associate Professor, Faculty of Science and Technology, American International University Bangladesh (AIUB) for his comments which were instrumental in the implementation phase of the thesis.

4 | P a g e

# Abstract

Pathological specimens are scanned using a whole slide imaging (WSI) scanner to create digital images for monitor-based diagnosis and analysis. The slide needs to be rescanned if the image quality is occasionally unsatisfactory because of focus-error or noise. An evaluation method for reference less quality was previously proposed, although some artifacts (such as tissue folds and air bubbles) were identified as false positives. Those artifacts must be disregarded when deciding whether or not rescanning is required because slide preparation, not scanning, is what causes them. By differentiating between the sources of quality degradation—the focus-error or noise created by the scanner and the artifact occurred during the slide preparation—this research suggests a strategy for a more practical approach to assess WSI quality. In the method, two UNet based architecture: DoubleUNet and ResUNet++ is used to segment the area in which the artifact is located. Besides the segmentation, the artifact severity is also analyzed using ensemble learning. To do the ensemble learning, at first transfer learning was implemented using various pre-trained model and then, from the generated model best models are selected for ensemble learning. In case of artifact segmentation above 97% accuracy is achieved and in the case of severity analysis 99.99% accuracy is gained which is almost perfect.

# Table of Contents

# List of Figures

# List of Tables

# 1.Introduction

The development of a whole slide imaging (WSI) scanner made it possible to digitize pathological works [1][2][3]. Glass slide specimens are converted into high-resolution digital images by the scanner for use in computational pathology and pathology practice. In hospitals and pathology labs, it enables us to undertake picture analysis and diagnosis using automated algorithms or human viewing on a computer screen. It pushes the boundaries of light microscopy and has great promise not just for research and education but also for primary and secondary diagnoses [4]. However, it is important to take into account a number of factors before using a WSI scanner as a standard medical device, including image quality, color variability, inadequate consistency of slide preparation, and image format [3][5]. In this research, we set our focus on mainly tissue artifact area segmentation and artifact severity analysis which are primarily base on image quality.

In research labs, hospitals, medical colleges and pathology departments a whole Slide Imaging System would be used to scan biological specimens on a regular basis [6]. WSI scanners may produce images of insufficient quality. Images of poor quality can seriously hinder diagnosis and render analysis ineffective. The operator is currently required by the WSI system to guarantee the accuracy of the scanned image. However, such manual evaluation is expensive and unreliable since it is susceptible to fatigue, inconsistent results, and prejudice. In order to provide access to high-quality photographs, it is important to analyze the quality of scanned images automatically [7].

Failures during scanning might result in focus blur or noise being introduced into the WSI scanner, which is one of the main reasons of image quality loss. For scanning the glass slides, focus points can be manually or automatically chosen. When focus points are chosen from an area with a different focus depth than the normal tissue area, the surrounding area is rendered out of focus. The independent random value that differs from adjacent image pixels is known as noise. Figure 1 demonstrates the noise, focus blur and under graded tissue area.

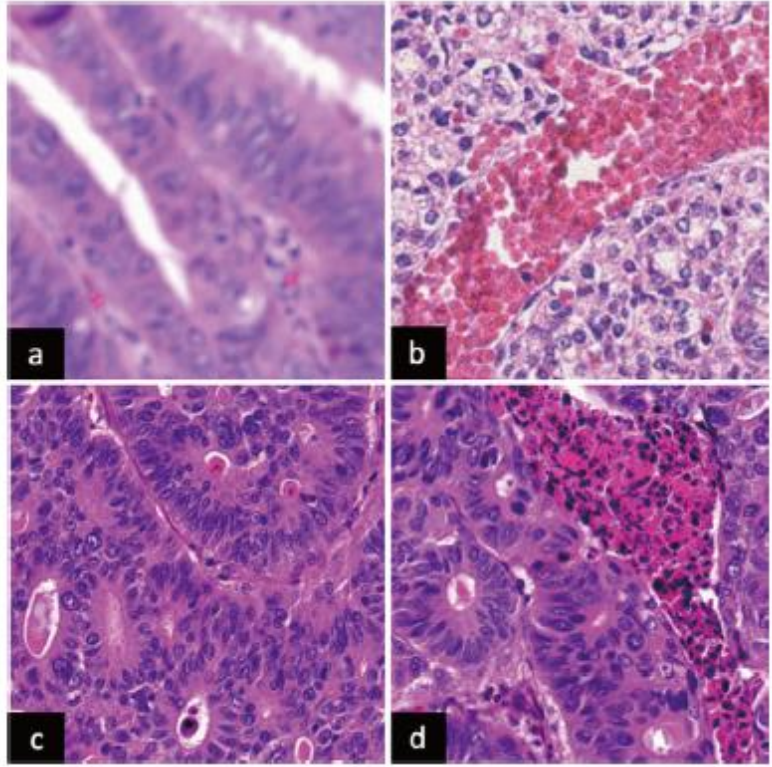

*Figure 1: (a) focus blur affected (b) noisy (c, d) under graded region of interest [7].*

It is important to rescan the slide if the focus error or noise significantly affects the scanned image. Hashimoto et al. previously suggested a reference less quality evaluation approach (RQM) to assess the scanned image's quality in order to identify slides that needed to be rescanned. The major causes of quality failure in the WSI scanner are sharpness and noise measurement, which is how this approach assesses the quality of images.

Tissue artifacts are produced during the glass slide preparation stage, which is another reason why image quality fails [9]. The two main artifacts discovered in WSI are a tissue fold and an air bubble [10], as illustrated in fig. 2. Tissue artifacts deceive analysis and diagnosis by concealing or changing important information. Furthermore, focusing errors are brought on by such objects [11]. They present various focus depths, and if focus locations are chosen from these artifacts, the surrounding area is rendered out of focus. Tissue artifacts like air bubbles and tissue fold would also be found using the prior RQM approach [8]. If a slide has a scanning problem, rescanning it can restore the image's quality but not remove tissue artifacts. It is useless to rescan the slide if the WSI quality is low due simply to tissue artifacts. For both analysis and quality assessment, artifacts should be found and disregarded. Therefore, to apply the judgment's outcome in the workflow and address the quality failure, an efficient system must identify the reason of quality failure.

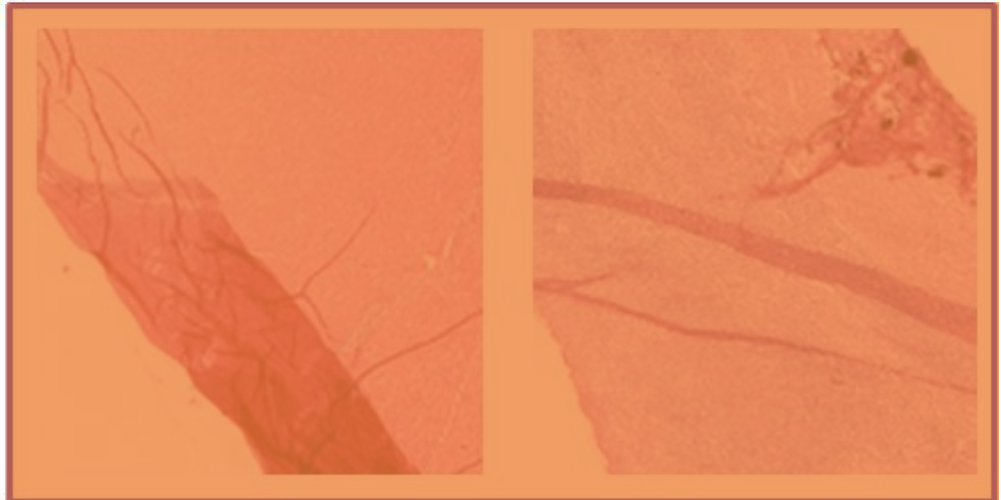
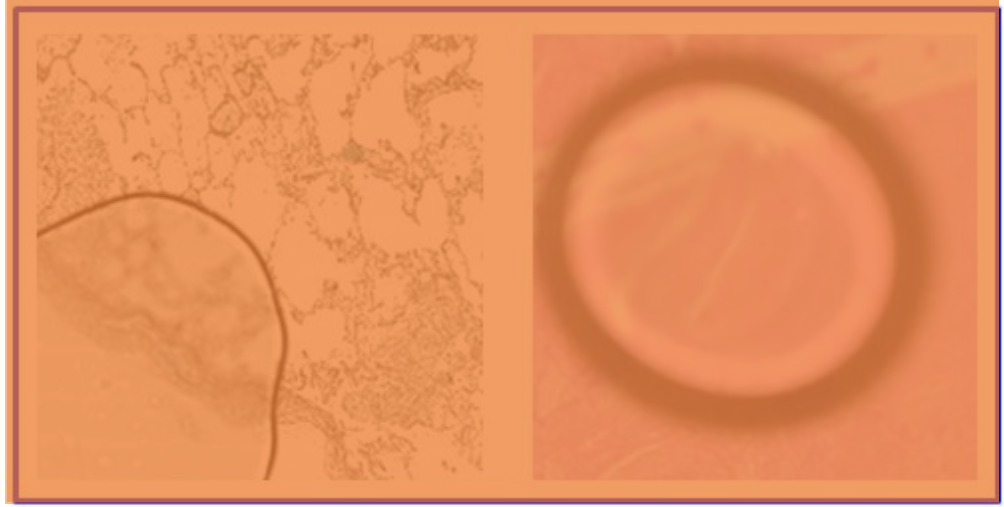

*Figure 2: Tissue artifacts (Tissue-fold and Air-bubble).*

But, simply removing all the artifacts is not a feasible solution. There can be such slides those contain minor artifacts while contain major tissue specimen which can be proven important to identify disease. So, those kind slides can generate important features for machine learning. But, due to presence of minor artifacts if those images are disregarded then again the same problem which is wrong diagnosis of disease can occur. That is why the severity of the artifact is to be determined before removing from the diagnosis process.

In this research, we suggest adding the tissue artifact detection stage and the severity analysis of the artifacts, as shown in fig. 3. The suggested method uses a UNet-based image segmentation methodology to identify artifacts, then classifies the images based on their severity before deciding whether or not to exclude them from the diagnosis process.

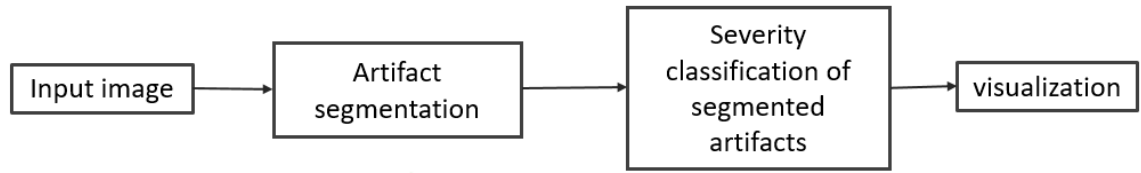

*Figure 3: Proposed method.*

## 2.Literature Review

Although reduced-reference methods are available in some situations, methods for evaluating image quality can be roughly categorized into two groups: those that use reference images and those that do not. However, because a perfect reference picture is not available, reference-based evaluation cannot be used to identify a quality problem in the WSI scanner system. For the purpose of evaluating the quality of applications involving cameras, displays, video transmission, and codecs, references-less quality evaluation methods (RQMs) have been created [12][13][14][15][16][17][18][19][20].

Focus inaccuracy and noise are the key problems in WSI applications for managing the quality of digital pathology [8][21][22][23]. The RQMs have been used in this way, taking specific distortion into account. The method put out by Hashimoto et al. based on sharpness and noise assessment predicts both a subjective quality score as well as an objective score, such as a root mean square error, because computational image analysis is a crucial issue in both visual diagnostics and digital pathology. In order to gauge the reproducibility of the scanner, Shrestha et al. provided a method to assess the WSI's quality [21]. This approach, however, is unsuitable for practical applications because it depends on a reference image for evaluation. Another technique used deep learning to find out-of-focus areas, but it ignored noise-induced distortion [22]. Because it is effective and simple, the system suggested in this work is based on the method developed by Hashimoto et al. To distinguish between the tissue artifacts and scanning errors, it is vital to take into account the source of degradation while using the WSI scanning procedure in digital pathology.

In several areas of computer vision, object detection or categorization in an image is carried out. Both supervised and unsupervised methods exist. The supervised method works well for the proposed system's air-bubble and tissue-fold detection because the unsupervised method could mix up artifacts with other anomalies that could have clinical significance. Template matching is a well-known and established supervised method for object detection. Statistical techniques like SVM and neural network classifiers have also been frequently used to identify objects in an image. Handcrafted features have been employed in this type of supervised classification, including local binary patterns, the gray level co-occurrence matrix (GLCM), and scale-invariant feature transform. It is well known that employing the sequential feature selection strategy when choosing the features for a classifier can increase performance [24][25].

More recently, non-handcrafted features that CNN gathered from the data were made possible by the development of deep learning technology. CNN can also be applied directly to an image. It has been demonstrated that a deep learning strategy, like CNN, performs better than a handmade feature based approach, while requiring a larger training dataset. When applied to a small dataset, it frequently becomes over fitted [26]. Transfer learning is typically employed in such instances to alleviate the problem of data limitation [27][28]. A network can avoid overfitting by using dropout and data augmentation [29][30]. In this work, we compared the segmentation method based on DoubleUNet and ResUNet++ with a VGG19 model as decoder. As for image classification process for severity analysis, we used various transfer learning methods for base model selection and lastly used several meta learners for stacking based ensemble learning.

In histopathology, object detection using segmentation and image classification methods were used in previous researches. A review paper on importance of artifact detection and possibility of wrong disease diagnosis due to presence of artifact is published by Syed Ahmed Taqi at el. in the year 2018 [31]. In the year 2019, Kay R.J. Oskal at el. Published a paper regarding a UNet based tissue segmentation method in WSI system [32]. An automated blur detection method is proposed by Xavier et al. in the year 2013 [33]. Samar khan published a paper regarding the probability of wrong diagnosis due to artifact in the year 2014 [34]. Ozan Oktay et al. introduced a unique attention gate (AG) model for medical imaging in 2018 that automatically learns to focus on target structures of various sizes and forms. Models trained with AGs intuitively learn to emphasize prominent features that are helpful for a particular task while suppressing irrelevant regions in an input image. Because of this, they can do away with the need for explicit external tissue/organ identification modules in cascaded convolutional neural networks (CNNs) [35]. A SVM based approach was taken by Hossain Md. Shakhawat in the year 2020 to evaluate image quality in WSI systems [7]. In the year 2019, Morteza Babaie et al. proposed a unique approach that mimerges the svm with dense layer 201 model to enhance the feature extraction process for medical images [36]. In the year, 2009, Pinky A. Bautista had done a research regarding artifact detection in WSI systems using basic object detection model [37]. In the year 2015, Hang Wu proposed a method regarding detecting blur artifacts in WSI [38]. Sonal Kothari published a research article on eliminating tissue fold artifact in the year 2013 [39]. Wei Lu et al. in the year 2013 researched on the presence of air bubble artifact in slides [40]. Pinky A. Bautista again in the year 2010 published another paper regarding enhancing the visualization and detection of tissue fold artifacts in whole slide imaging system [41].

So, from the literature, it is apparent that, continuous researches are being done by the researchers to further improve the accuracy of the detection of artifacts and also how to eliminate the artifacts automatically using different approaches.

## Problem Statement

Existing methods only detects the tissue artifacts but does not analyze the severity of a particular artifact.

But, severity analysis is important otherwise, significant amount of tissue can be eliminated from the specimen.

Till now no work has been done to analyze the severity of the artifact. That's why we focused our research on mainly two things:

- **Improve the accuracy of artifact segmentation.**
- **Analyze the artifact severity.**

# 3.Methodology

The WSI scanner is used for scanning whole slide images. It is hard to determine the region of interest using human eye. So various approaches are being taken to enhance the automatic artifact detection. Among those approaches there are several machine learning approaches applied for advanced image processing.

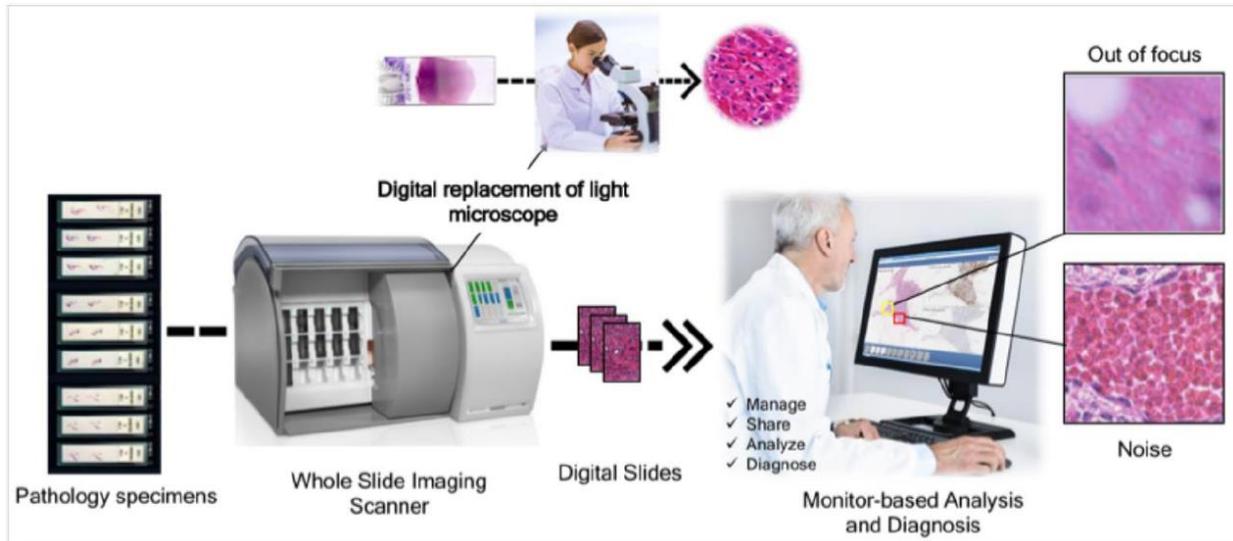

*Figure 4: WSI system [42].*

Pathological specimens are a bit complex to extract meaningful image data from. Because from a lot of images which are apparently similar, important distinguishable features have to be extracted. For this purpose, we have proposed a method which will extract necessary information which will be used for successful feature extraction; ultimately leading to efficient pathological analysis and disease diagnosis. In our proposed method, we have applied image annotation, data pre-processing, segmentation, classification and visualization respectively in an organized manner.

## 3.1. Overview of The Proposed Method

In the preliminary stage, the images are captured using WSI system. The whole image will be divided into many small parts and segmentation is done for each part.

The proposed method takes input images then detect the artifact using UNet based image segmentation approach. This segmentation process will identify the tissue artifact and will highlight that part while separating it from the background. After the segmentation process, the artifacts are being classified in terms of severity. For this classification task, Ensemble learning is applied. Ensemble learning requires several methods. For selecting those methods transfer learning is applied using different combinations of parameters. After the severity classification being done, the final output is generated and sent for visualization.

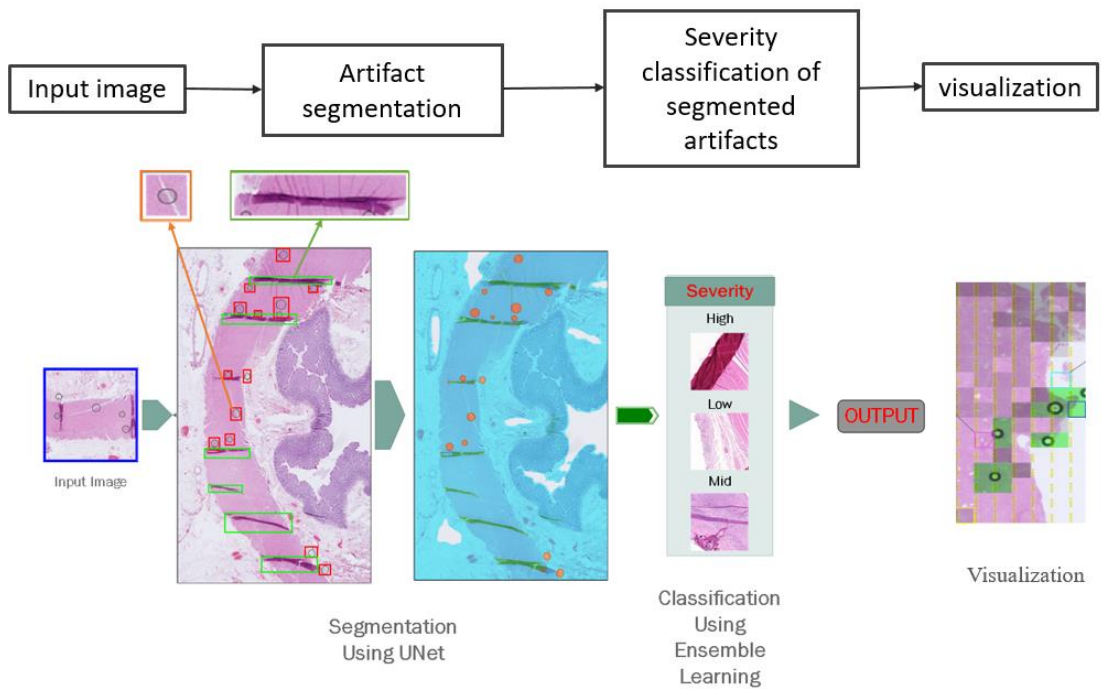

*Figure 5: Proposed method for artifact segmentation and severity analysis.*

In the visualization screen, those identified artifacts along with their severity level will be shown and pathologists will be able to successfully pinpoint the affected region.

## 3.2. Artifact Detection Method

For, artifact detection we have used UNet based architecture for image segmentation. At first we used basic UNet. But the results were not good. Then we used 'DoubleUNet' and 'ResUNet++' as they have become quite popular in recent times for medical image segmentation tasks.

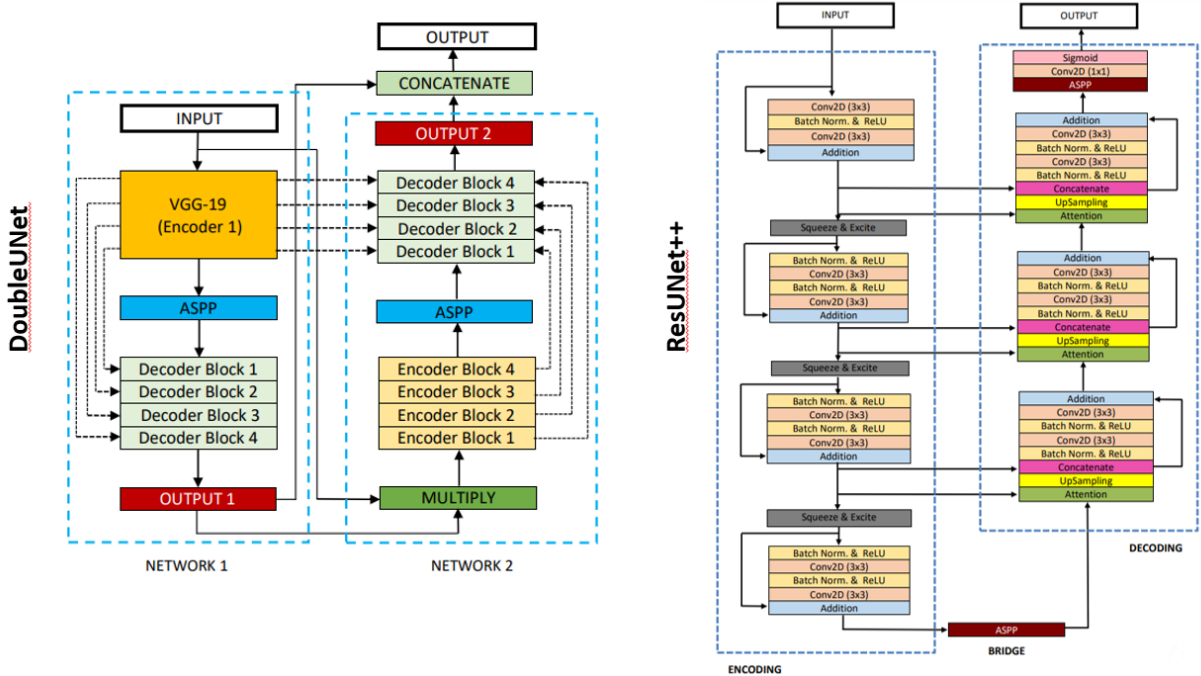

*Figure 6: DoubleUNet and ResUNet++ architecture.*

DoubleUnet architecture uses VGG-19 model as encoder. The first layer of UNet generates an output which is multiplied with the primary input and that multiplied output is given as input to the second layer of UNet. The primary output and the secondary output concatenate and generates the final output.

The ResUNet++ architecture follows a basic ResNet architecture and builds a bridge between the ResNet and UNet architecture. In simple words, it is a UNet architecture which follows ResNet architecture for segmentation.

## 3.3. Artifact Severity Classification

After the artifacts are being identified using segmentation techniques, the images need to be classified. For classification, we primarily used transfer learning using various pre-trained model. After getting the results we have compared the results according to the accuracy and loss values. Then we selected the best models to ensemble the results using stacking method.

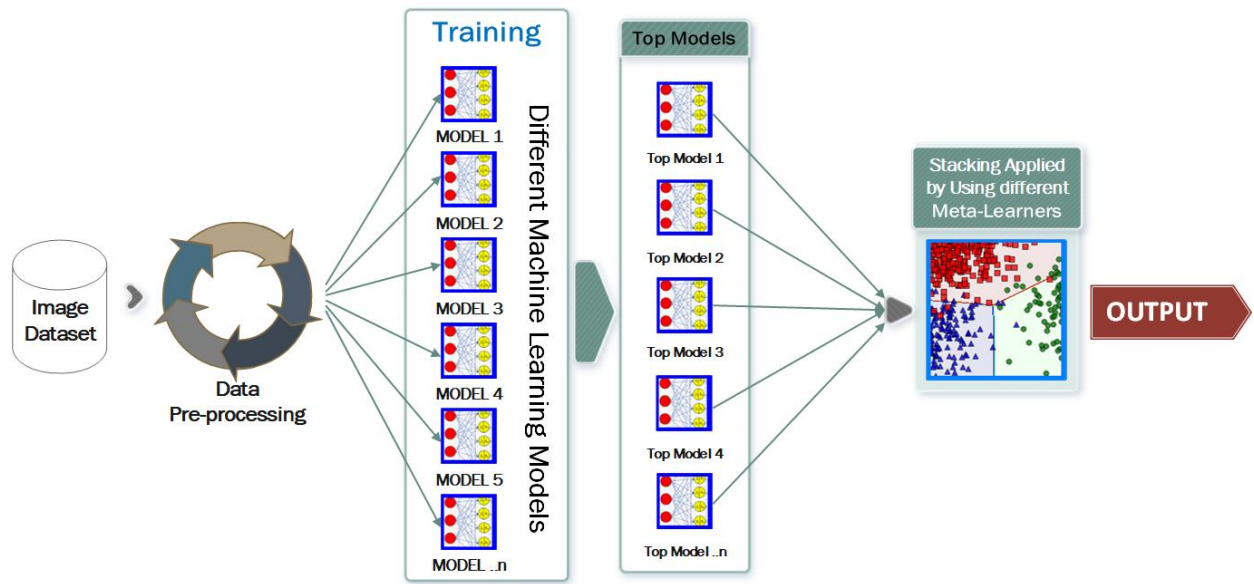

*Figure 7:* *Ensemble learning.*

For implementing ensemble learning, we tested 10 different meta learners:

- Logistic Regression
- K Nearest Neighbor
- Support Vector Machine
- Decision Tree
- Random Forest
- Ada Boost Classifier
- XGB Classifier
- Gradient Boost Regressor
- Gradient Boost Classifier
- Gaussian Naïve Bias

Severity analysis requires extensively complex calculation process for selecting differentiable features for image classification. The transfer learning models we used are not able to give perfect results from every perspective. Each model focuses on particular type of feature extraction policy. That's why, to combine those strategies, ensemble learning is used.

Within the ensemble learning itself there are approaches those are specific to different areas. Meta learners used in stacking policy acts in different ways. That's why several meta learners assessed to find out which meta learner generates the best results.

# 4.Experimment

## 4.1. Dataset

In total 26 slides are used in this experiment, which included the tissues of different organs including kidney, liver, heart, brain etc. So the dataset is organ independent.

The slides were collected form the several biomedical laboratories of different parts of the world.

**Segmentation task**

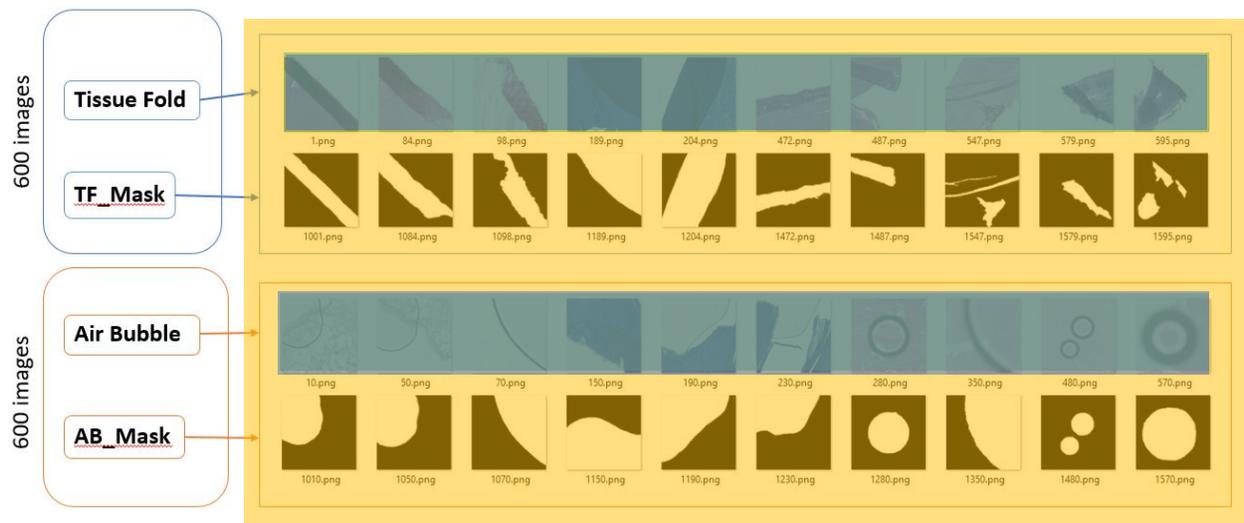

Figure 8: *Dataset for segmentation.*

Each category consists of 600 images. Data augmentation was applied for data balancing. Every images were segmented manually in adobe Photoshop by using binary masking function.

**Classification task**

Taking images from the same source the dataset was categorized in three categories:

- **High severity**
- **Low severity**
- **Mid severity**

The dataset was again augmented for balancing purpose. This time the dataset contains 1380 images. This data annotation was done by certified expert.

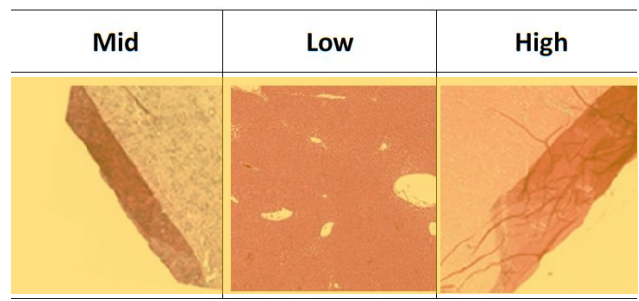

*Figure 9: Dataset for severity classification.*

## 4.2. Experimental Setup

Our experiment is divided in two parts: a) artifact segmentation; b) severity analysis/ classification.

### 4.2.1. Experimental setup for artifact segmentation

For the segmentation experiment the hyper parameters were tuned in different combinations.

Table 1: *Hyper parameter tuning for segmentation experiment.*

| | |
|---|---|
| Train images | 480 |
| Valid images | 60 |
| Test images | 60 |
| Batch size | 8 |
| UNet architecture | **DoubleUNet,** ResUNet++ |
| Optimizer | Adam, Adamax, **RMSprop**, SGD |
| Loss function | **Dice Coef Loss**, binary cross entropy |
| Learning Rate | Starting from **1e$^{-4}$** |
| ReduceLROnPlateau | **monitor='val_loss', factor=0.1, patience=4** |
| EarlyStopping | **monitor='val_loss', patience=10** |

Different epochs were tried but the result was not always uniformly tuned and that's why we used early stopping function to prevent overfitting and gain a reliable value for segmentation.

### 4.2.2. Experimental setup for severity classification

For classification task we used the following combinations described in table 2 and table 3.

*Table 2: Hyper parameter optimization.*

| Hyper parameter | Optimization Space |
|---|---|
| Epoch | 25 |
| Batch size | 32 |
| Learning rate | 0.0001 |
| Optimizer | 'Adam', 'Adamax', 'RMSprop' |
| Loss function | [Categorical cross entropy], [Kullback Leibler Divergence] |
| Class mode | categorical |

For implementing ensemble learning, we had to run transfer learning several times using different pre-trained model in various combinations. We used the most famous 10 pre-trained model from the official website of 'keras'.

*Table 3: Pre-trained models; their image dimensions.*

| Data Augmentation | | | Image input dimension | |
|---|---|---|---|---|
| **Parameter** | **Value** | | Xception | 224×224 |
| Rescale | 1./255 | | VGG16 | 224×224 |
| Zoom range | 0.3 | | VGG19 | 224×224 |
| Rotation range | 15. | | ResNet50 | 224×224 |
| Horizontal flip | True | | InceptionV3 | 224×224 |
| **Data set** | | | InceptionResNetV2 | 224×224 |
| **Total images** | **Train images** | **Test images** | MobileNet | 224×224 |
| High | 320 | 140 | MobileNetV2 | 224×224 |
| Low | 320 | 140 | DenseNet121 | 224×224 |
| Mid | 320 | 140 | NasNetLarge | 331×331 |
| Grand Total | **960** | **420** | | |
| The data set contains total **1380** images. The train, test ratio is 70% : 30% | | | | |

**Total 60 combinations of code were run to complete the experiment.**

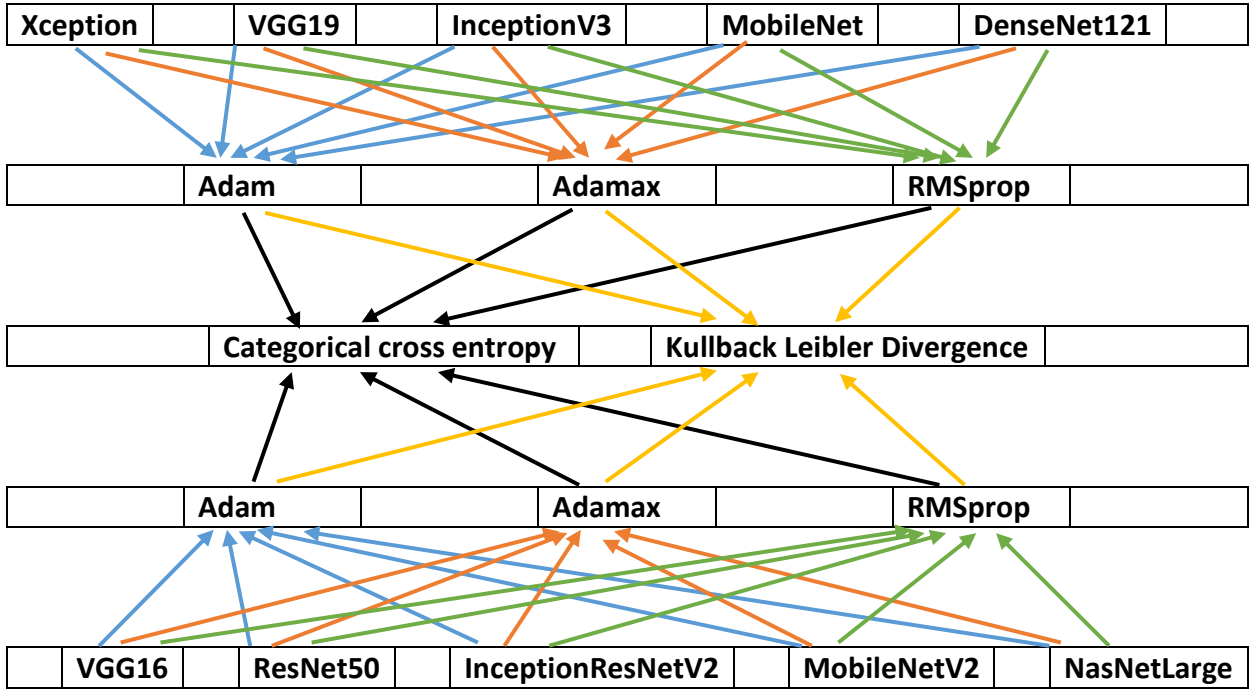

| Pretrained Model | Optimizer | Loss function | calculation | Total combination |
|---|---|---|---|---|
| 10 | 3 | 2 | 10 × 3 × 2 | 60 |

***Note:** "Accuracy Graph, ROC Curve and Confusion Matrices are generated for every combination but, for the paper writing convenience only the Accuracy graph, ROC curve and Confusion matrices of the Top 6 Models are mentioned in the Result section".

# 5.Results and Discussion

## 5.1. Artifact Segmentation Result

**Tissue Fold: ResUNet++**

Table 4 represents the hyper-parameter combination for the best result for tissue fold artifact using ResUNet++ architecture. RMSprop optimizer performs better than any other optimizers. The learning rate was set to 0.0001. But, the 'ReduceLROnPlateau' function reduces the learning rate at every step where validation loss increases. The training was stopped when lr was **$1e^{-9}$**.

*Table 4: Best segmentation result for Tissue-fold using ResUNet++.*

| Train images | 480 |
|---|---|
| Valid images | 60 |
| Test images | 60 |
| Batch size | 8 |
| Epoch | **38 Epoch** |
| Optimizer | **RMSprop** |
| Loss function | **Dice Coef Loss** |
| Learning Rate | **$1e^{-4}$** > $1e^{-5}$ > $1e^{-6}$ > $1e^{-7}$ > $1e^{-8}$ > **$1e^{-9}$** |
| ReduceLROnPlateau | **monitor='val_loss', factor=0.1, patience=4** |
| EarlyStopping | **monitor='val_loss', patience=10** |

| **loss** | **-0.9736** | **Val_loss** | **-0.9791** |
|---|---|---|---|
| **recall** | **29.65%** | **Val_recall** | **34.97%** |
| **precision** | **99.99%** | **Val_ precision** | **99.97%** |
| **Mean-IOU** | **85.81%** | **Val_mean-IOU** | **88.88%** |
| **IOU** | **94.87%** | **Val_ IOU** | **96.05%** |
| **Dice Coef** | **97.36%** | **Val_Dice Coef** | **97.98%** |

Dice coef loss is defined as negative value that's why the loss values are negative.

**Graph**

Figure 10 represents the graph of mean IOU, Loss curve, precision curve, recall curve, dice-coefficient and the IOU. The last two graphs which are the dice-coefficient and IOU curve defines the actual performance of the model.

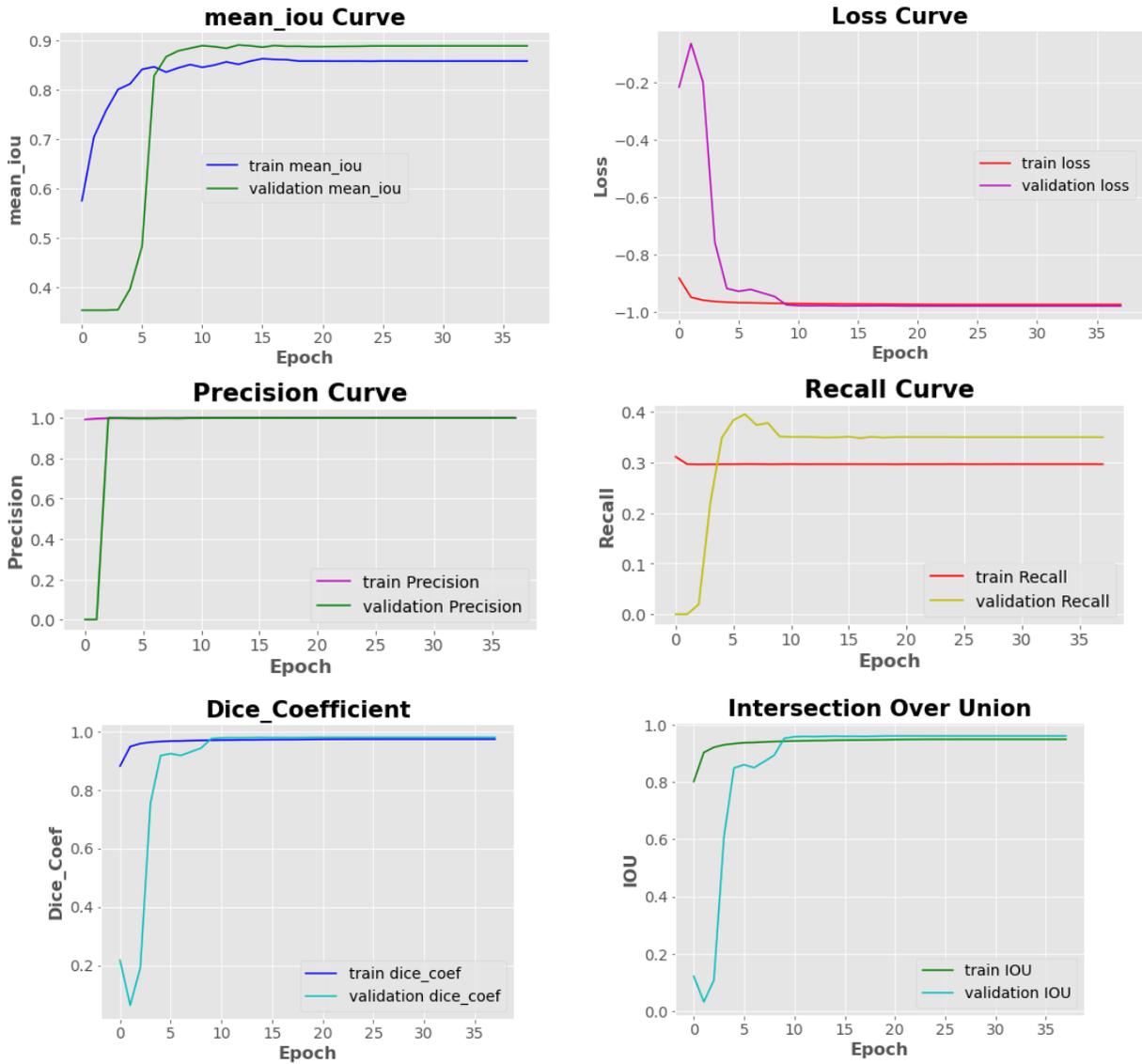

*Figure 10: Best results of ResUNet++ on Tissue-fold dataset.*

**Tissue Fold: DoubleUNet**

Table 5 represents the hyper-parameter combination for the best result for tissue fold artifact using DoubleUNet++ architecture. RMSprop optimizer performs better than any other optimizers. The learning rate was set to 0.0001 and was consistent throughout the whole training process. The result was achieved at epoch 50.

*Table 5: Best segmentation result for Tissue-fold using DoubleUNet.*

| Train images | 480 |
|---|---|
| Valid images | 60 |
| Test images | 60 |
| Batch size | 8 |
| Epoch | **50 EPOCH** |
| Optimizer | **RMSprop** |
| Loss function | **Dice Coef Loss** |
| Learning Rate | **1e$^{-4}$** |
| ReduceLROnPlateau | **monitor='val_loss', factor=0.1, patience=10** |
| EarlyStopping | **monitor='val_loss', patience=10** |

| **loss** | **-0.9692** | **Val_loss** | **-0.9762** |
|---|---|---|---|
| **recall** | **29.64%** | **Val_recall** | **35.26%** |
| **precision** | **99.98%** | **Val_ precision** | **99.97%** |
| **mean-IOU** | **41.03%** | **Val_mean-IOU** | **35.12%** |
| **IOU** | **94. 03%** | **Val_ IOU** | **95.50%** |
| **Dice Coef** | **96.92%** | **Val_Dice Coef** | **97.69%** |

Dice coef loss is defined as negative value that's why the loss values are negative.

**Graph**

Figure 12 represents the graph of mean IOU, Loss curve, precision curve, recall curve, dice-coefficient and the IOU. The last two graphs which are the dice-coefficient and IOU curve defines the actual performance of the model.

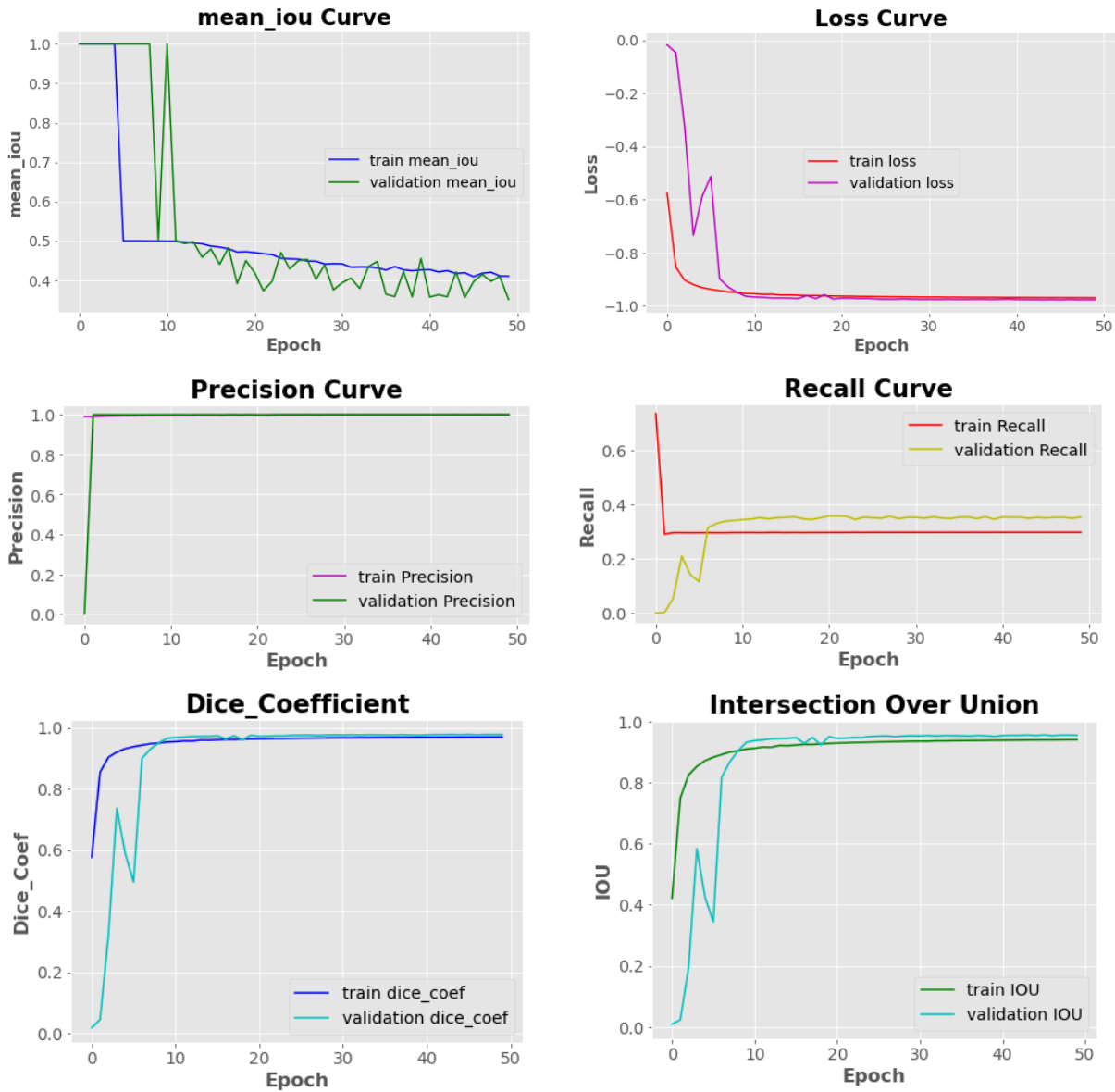

*Figure 11: Best results of DoubleUNet on Tissue-fold dataset.*

## Overall comparison of Tissue-Fold detection between ResUNet++ and DoubleUNet

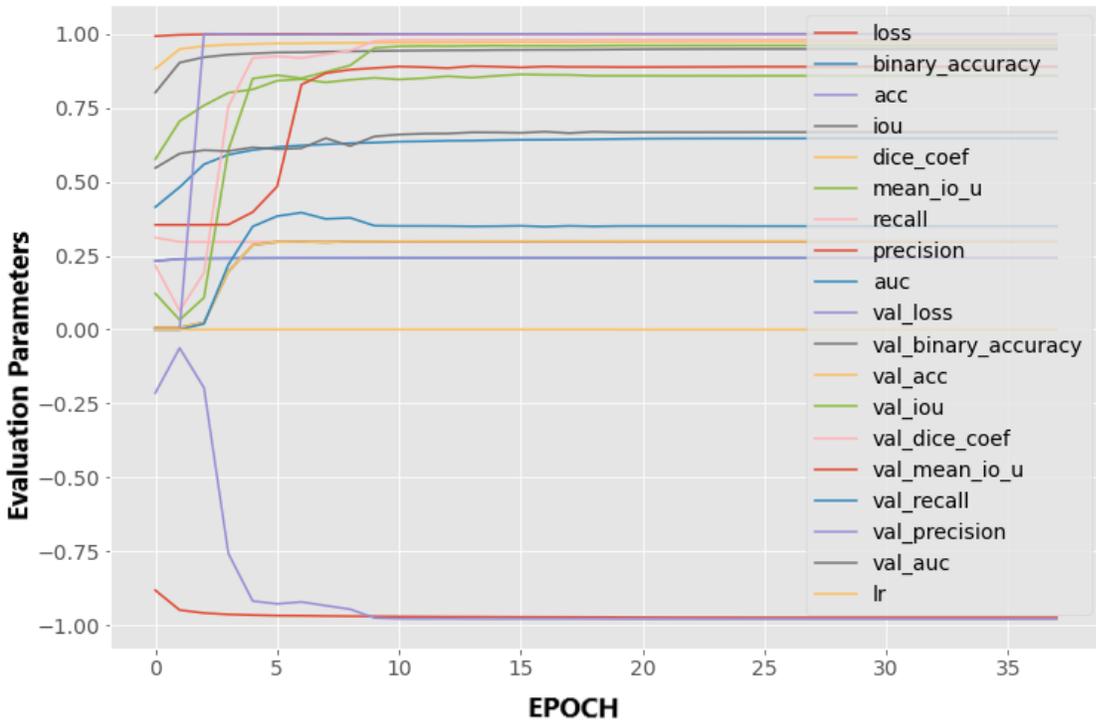

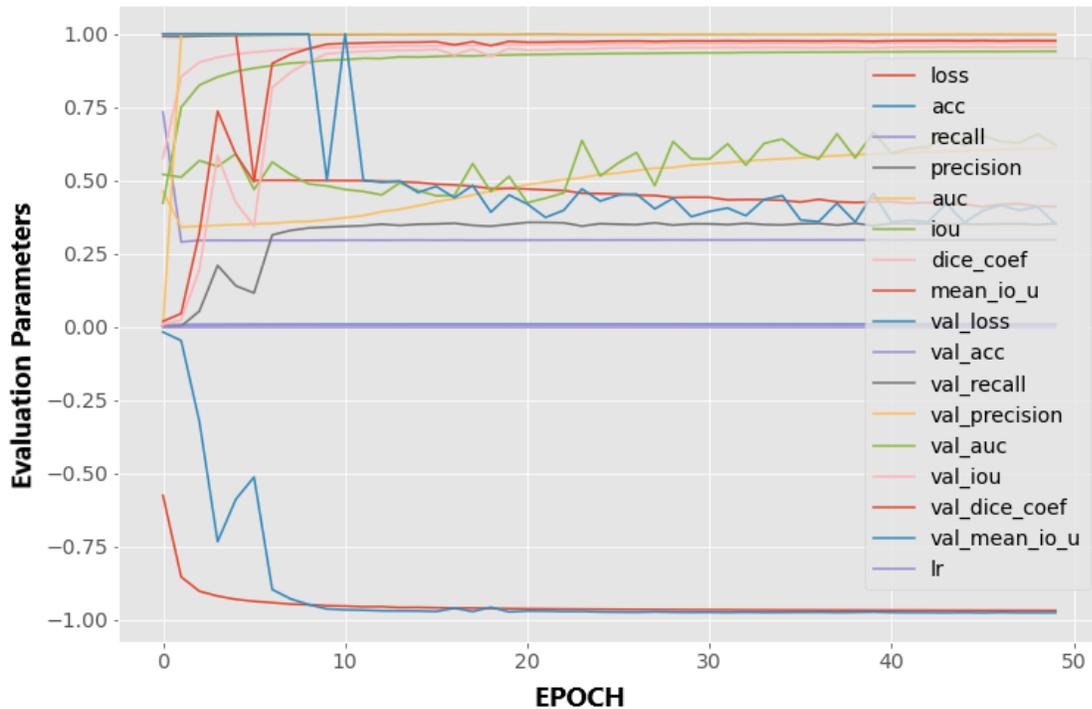

*Figure 12: Comparison graph for ResUnet++ and DoubleUNet for Tissue-fold dataset.*

## Air Bubble: ResUNet++

Table 6 represents the hyper-parameter combination for the best result for air bubble artifact using ResUNet++ architecture. RMSprop optimizer performs better than any other optimizers. The learning rate was set to 0.0001. But, the 'ReduceLROnPlateau' function reduces the learning rate at every step where validation loss increases. The training was stopped when lr was **$1e^{-13}$**.

*Table 6: Best segmentation result for Air-bubble using ResUNet++.*

| Train images | 480 |
|---|---|
| Valid images | 60 |
| Test images | 60 |
| Batch size | 8 |
| Epoch | **56 EPOCH** |
| Optimizer | **RMSprop** |
| Loss function | **Dice Coef Loss**, binary cross entropy |
| Learning Rate | $1e^{-4} > 1e^{-5} > 1e^{-6} > 1e^{-7} > 1e^{-8} > 1e^{-9} > 1e^{-10} > 1e^{-11} > \mathbf{1e^{-12}}$ |
| ReduceLROnPlateau | **monitor='val_loss', factor=0.1, patience=4** |
| EarlyStopping | **monitor='val_loss', patience=10** |

| **loss** | **-0.9858** | **Val_loss** | **-0.9873** |
|---|---|---|---|
| **recall** | **40.62%** | **Val_recall** | **45.50%** |
| **precision** | **99.99%** | **Val_ precision** | **99.99%** |
| **mean-IOU** | **89.96%** | **Val_mean-IOU** | **91.32%** |
| **IOU** | **97.21%** | **Val_ IOU** | **97.45%** |
| **Dice Coef** | **98.58%** | **Val_Dice Coef** | **98.71%** |

Dice coef loss is defined as negative value that's why the loss values are negative.

**Graph**

Figure 15 represents the graph of mean IOU, Loss curve, precision curve, recall curve, dice-coefficient and the IOU. The last two graphs which are the dice-coefficient and IOU curve defines the actual performance of the model.

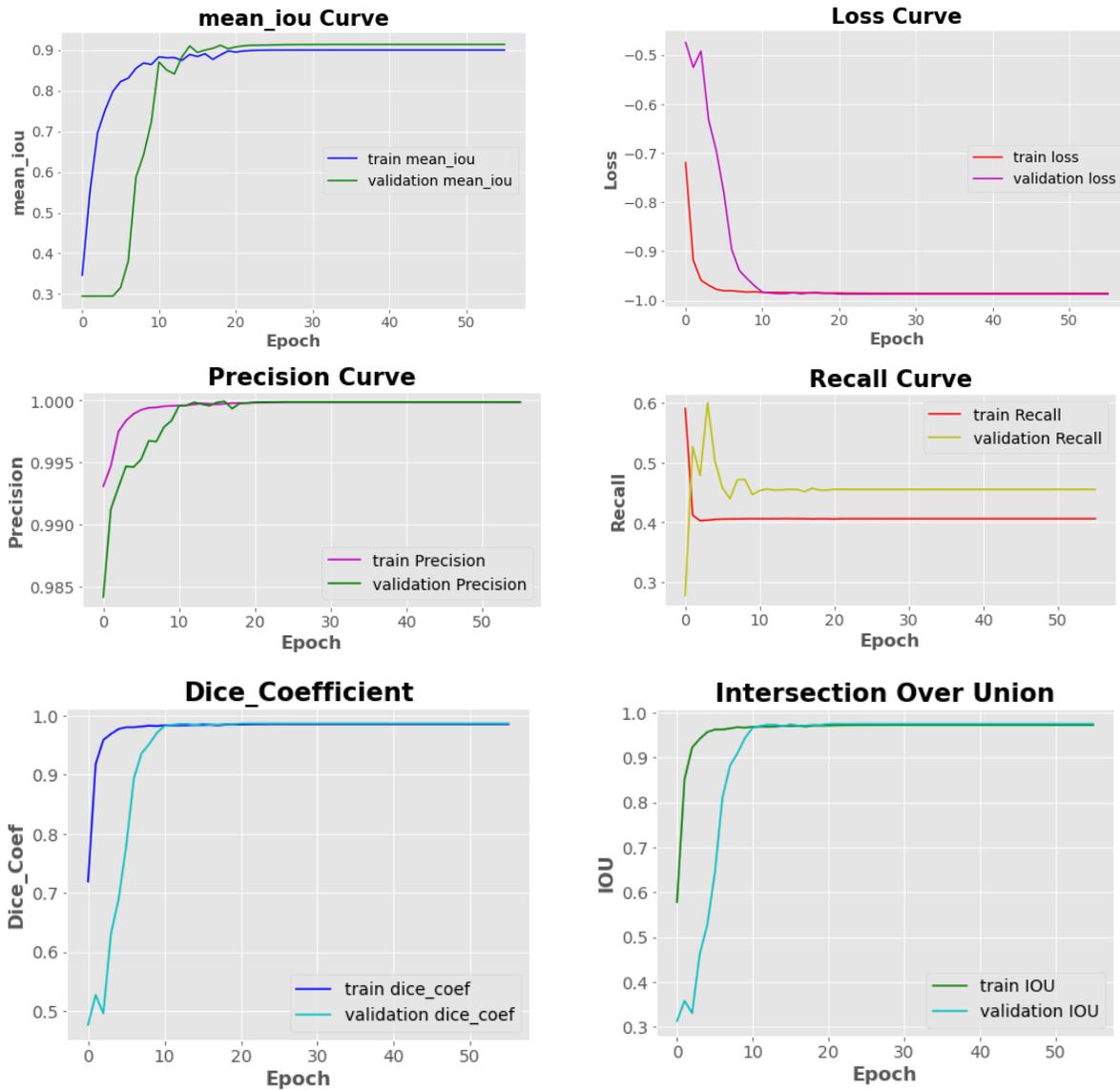

*Figure 13: Best results of ResUNet++ on Air-bubble dataset.*

## Air Bubble: DoubleUNet

Table 7 represents the hyper-parameter combination for the best result for tissue fold artifact using DoubleUNet++ architecture. RMSprop optimizer performs better than any other optimizers. The learning rate was set to 0.0001 and was consistent throughout the whole training process. The result was achieved at epoch 50.

*Table 7: Best segmentation result for Air-bubble using DoubleUNet.*

| Train images | 480 | | |
|---|---|---|---|
| Valid images | 60 | | |
| Test images | 60 | | |
| Batch size | 8 | | |
| Epoch | **50 EPOCH** | | |
| Optimizer | **RMSprop** | | |
| Loss function | **Dice Coef Loss** | | |
| Learning Rate | **1e$^{-4}$** | | |
| ReduceLROnPlateau | **monitor='val_loss', factor=0.1, patience=10** | | |
| EarlyStopping | **monitor='val_loss', patience=10** | | |
| **loss** | -0.9817 | **Val_loss** | -0.9841 |
| **recall** | 40.60% | **Val_recall** | 45.45% |
| **precision** | 99.97% | **Val_ precision** | 99.96% |
| **mean-IOU** | 32.97% | **Val_mean-IOU** | 30.82% |
| **IOU** | 96.41% | **Val_ IOU** | 96.83% |
| **Dice Coef** | 98.17% | **Val_Dice Coef** | 98.39% |

Dice coef loss is defined as negative value that's why the loss values are negative.

**Graph**

Figure 17 represents the graph of mean IOU, Loss curve, precision curve, recall curve, dice-coefficient and the IOU. The last two graphs which are the dice-coefficient and IOU curve defines the actual performance of the model.

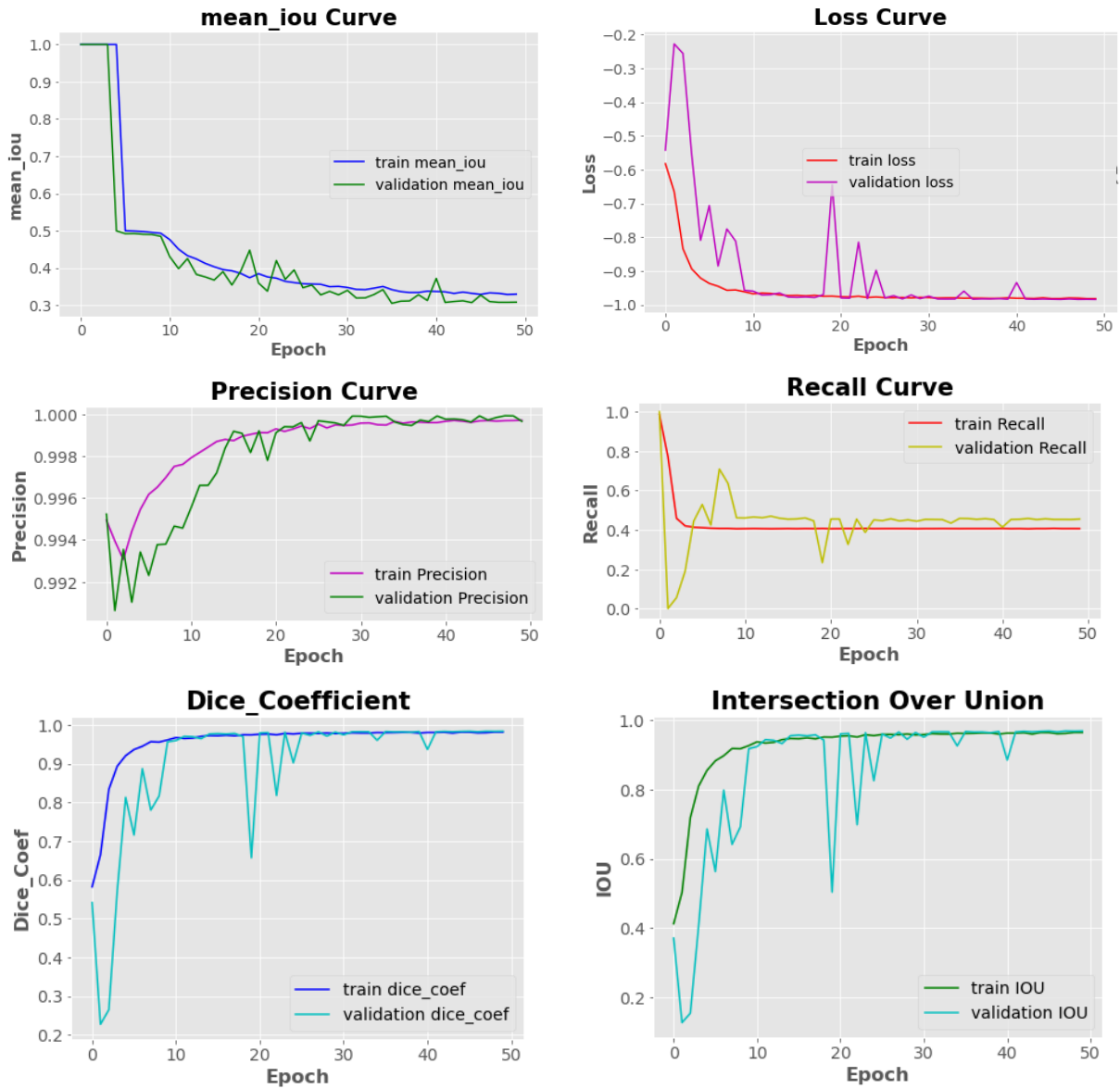

*Figure 14: Best results of DoubleUNet on Air-bubble dataset.*

**Overall comparison of Air-bubble detection between ResUNet++ and DoubleUNet**

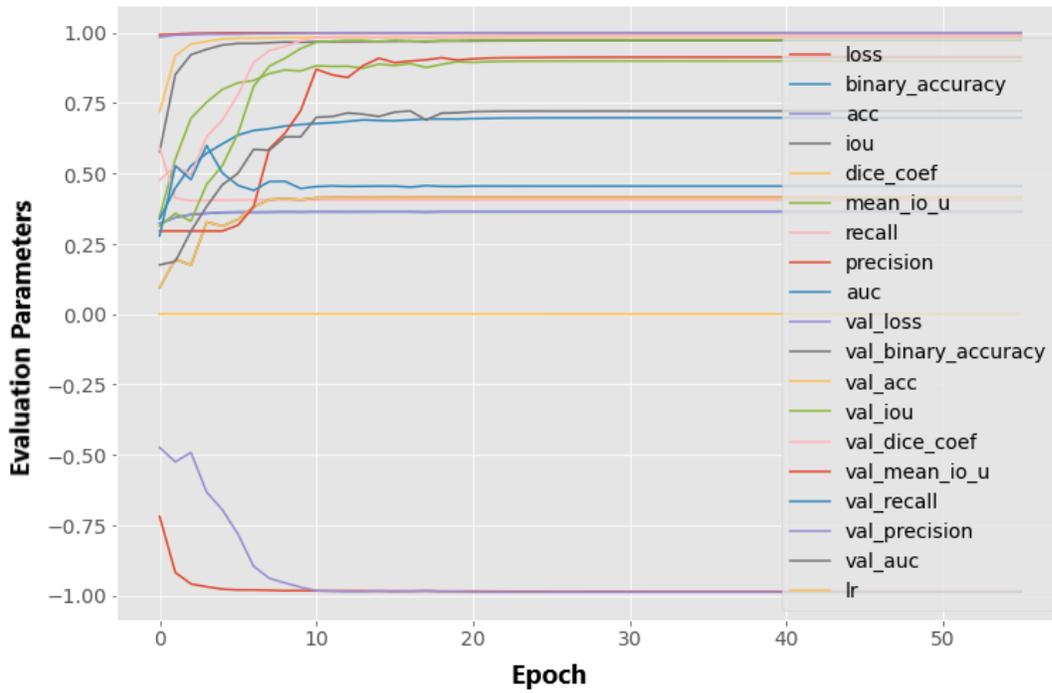

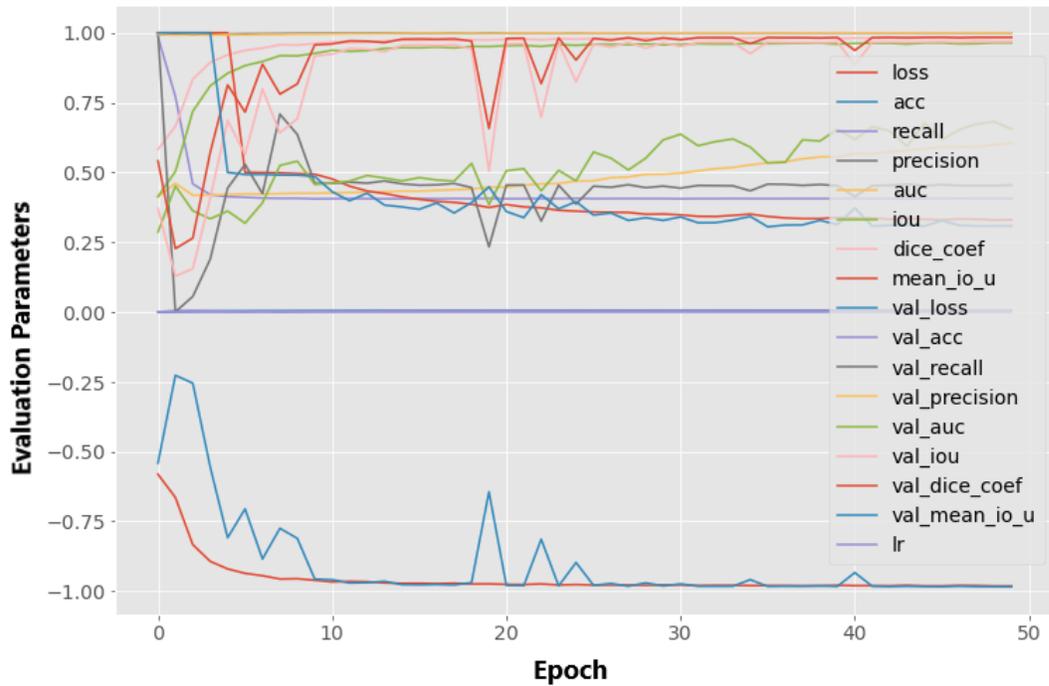

*Figure 15: Comparison graph of ResUNet++ and DoubleUNet for Air-bubble dataset.*

## Different Optimizers Tested for comparison

**"AVERAGE TEST IOU"** is calculated using the IOU value of the test set.

**Here, Test data set contains 60 images.**

$$*\text{AVG TEST IOU} = \frac{\text{IOU of 1st image + IOU of 2nd image + IOU of 3rd image + } \ldots \ldots + \text{IOU of 60th image}}{60}$$

**Test Accuracy is determined by using different IOU threshold values.**

**If Value > threshold, then append (1: True) If Value < threshold then append (0: False)**

**[ *For example, If, 58 out of 60 are true then test accuracy is (58/ 60) *100 %= 96.66 %]**

Table 8: Accuracy and IOU for Tissue Fold dataset.

| UNet | Optimizer | AVERAGE Test IOU | Test Accuracy (IOU Threshold = 90%) | Test Accuracy (IOU Threshold = 85%) |
|---|---|---|---|---|
| Double UNet | Adamax | 96.54 % | 90.00 % | 95.00 % |
| | Adam | 97.18 % | 95.00 % | 98.33 % |
| | RMSprop | 98.02 % | 98.33 % | 100 % |
| | SGD | 28.24 % | 0 .00% | 0.00 % |
| ResUNet ++ | Adamax | 95.74 % | 88.33 % | 96.66 % |
| | Adam | 96.95 % | 96.66 % | 100 % |
| | RMSprop | 97.72 % | 96.66 % | 100 % |
| | SGD | 79.62 % | 36.66 % | 50.00 % |

Table 9: Accuracy and IOU for Air Bubble dataset.

| UNet | Optimizer | AVERAGE Test IOU | Test Accuracy (IOU Threshold = 90%) | Test Accuracy (IOU Threshold = 85%) |
|---|---|---|---|---|
| Double UNet | Adamax | 99.08 % | 100 % | 100 % |
| | Adam | 91.35 % | 68.33 % | 81.66 % |
| | RMSprop | 99.11 % | 100 % | 100 % |
| | SGD | 45. 96 % | 0.00 % | 0.00 % |
| ResUNet ++ | Adamax | 97.43 % | 96.66 % | 98.33 % |
| | Adam | 98.80 % | 100 % | 100 % |
| | RMSprop | 98.15 % | 96.66 % | 96.66 % |
| | SGD | 28.29 % | 0.00 % | 0.00 % |

After doing all the combinations, it was found that "**RMSprop**" performs better than the other three optimizers. And SGD generates the poorest result.

So, in the next steps, for K-fold cross validation, "**RMSprop**" is used for both 'DoubleUNet' & 'ResUNet++'.

## 6-Fold Cross validation

Table 10: 6-Fold Cross validation for Tissue fold dataset.

|  | Fold no. | Training Set | Testing Set | Test Acc (IOU Threshold = 90%) |
|---|---|---|---|---|
| **DoubleUNet** | Fold 1 | A, B, C, D, E | F | 98.33 % |
|  | Fold 2 | B, C, D, E, F | A | 96.66 % |
|  | Fold 3 | C, D, E, F, A | B | 95.00 % |
|  | Fold 4 | D, E, F, A, B | C | 98.33 % |
|  | Fold 5 | E, F, A, B, C | D | 100 % |
|  | Fold 6 | F, A, B, C, D | E | 100 % |
|  | Average |  |  | **97.96%** |
| **ResUNet++** | Fold 1 | A, B, C, D, E | F | 95.00 % |
|  | Fold 2 | B, C, D, E, F | A | 96.66 % |
|  | Fold 3 | C, D, E, F, A | B | 96.66 % |
|  | Fold 4 | D, E, F, A, B | C | 95.00 % |
|  | Fold 5 | E, F, A, B, C | D | 98.33 % |
|  | Fold 6 | F, A, B, C, D | E | 96.66 % |
|  | Average |  |  | **96.38 %** |

Table 11: 6-Fold Cross validation for Air bubble dataset.

|  | Fold no. | Training Set | Testing Set | Test Acc (IOU Threshold = 90%) |
|---|---|---|---|---|
| **DoubleUNet** | Fold 1 | A, B, C, D, E | F | 100 % |
|  | Fold 2 | B, C, D, E, F | A | 100 % |
|  | Fold 3 | C, D, E, F, A | B | 98.33 % |
|  | Fold 4 | D, E, F, A, B | C | 100 % |
|  | Fold 5 | E, F, A, B, C | D | 100 % |
|  | Fold 6 | F, A, B, C, D | E | 100 % |
|  | Average |  |  | **99.72 %** |
| **ResUNet++** | Fold 1 | A, B, C, D, E | F | 96.66 % |
|  | Fold 2 | B, C, D, E, F | A | 100 % |
|  | Fold 3 | C, D, E, F, A | B | 100 % |
|  | Fold 4 | D, E, F, A, B | C | 98.33 % |
|  | Fold 5 | E, F, A, B, C | D | 100 % |
|  | Fold 6 | F, A, B, C, D | E | 100 % |
|  | Average |  |  | **99.16 %** |

## ROC Curve

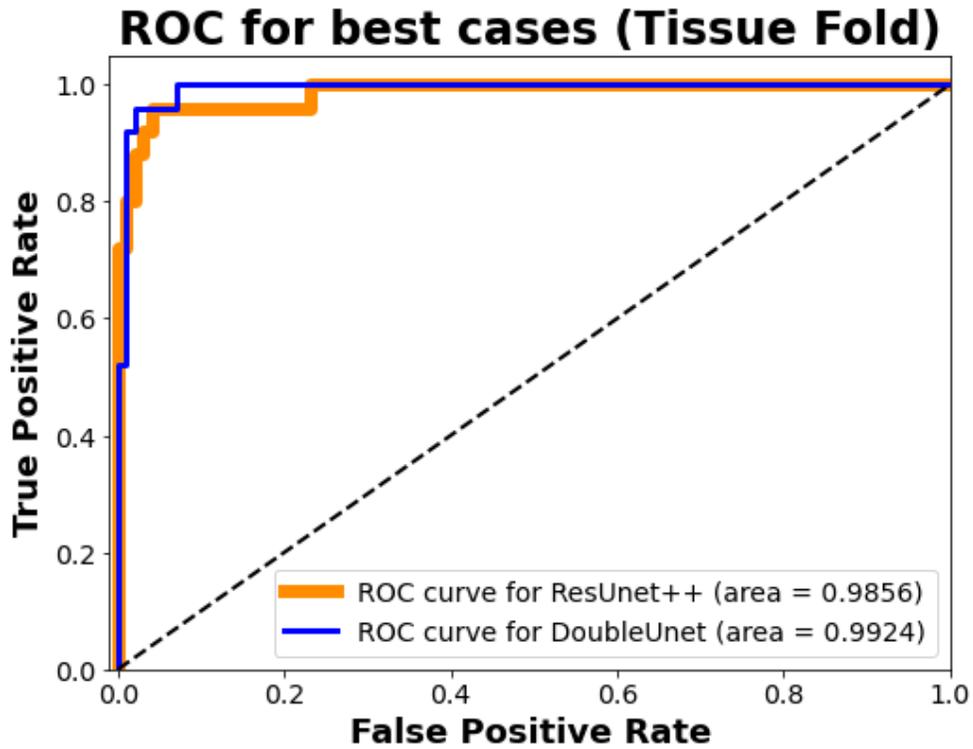

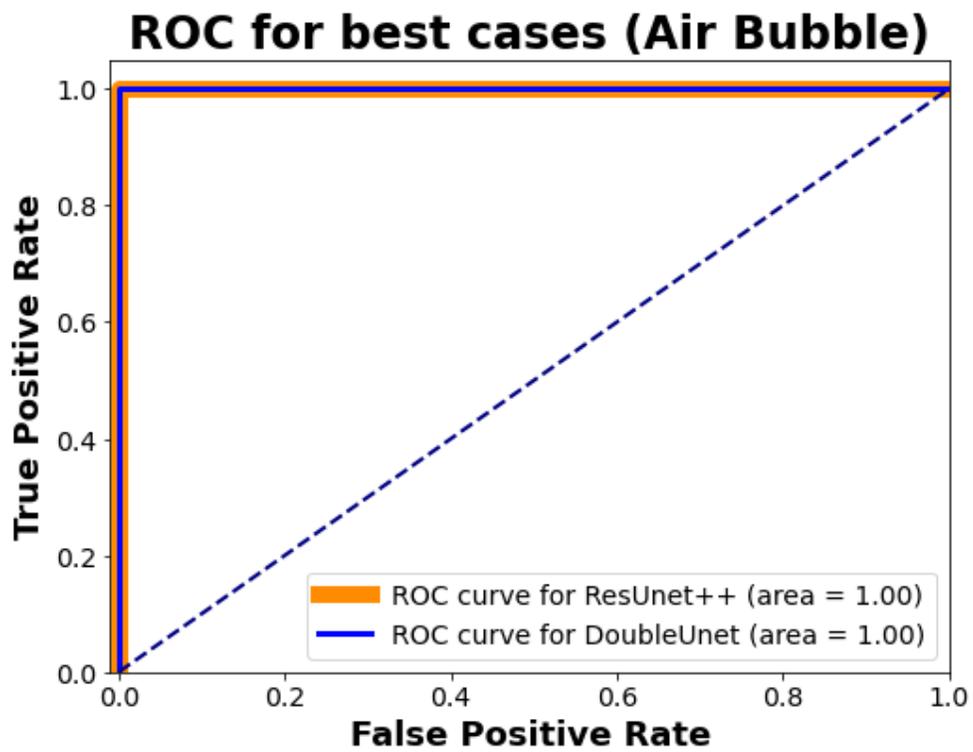

*Figure 16: ROC comparison.*

*Table 12: Over all comparison of segmentation models.*

| Artifact | UNet | Train Loss | Val Loss | Train Recall | Val Recall | Train Precision | Val Precision | Train IOU | Val IOU | Train Dice-Coef | Val Dice-Coef |
|---|---|---|---|---|---|---|---|---|---|---|---|
| Tissue Fold | Double UNet | -0.9692 | -0.9762 | 0.2964 | 0.3526 | 0.9998 | 0.9997 | 0.9403 | 0.9550 | 0.9692 | 0.9769 |
| Tissue Fold | ResUNet++ | -0.9736 | -0.9791 | 0.2965 | 0.3497 | 0.9999 | 0.9997 | 0.9487 | 0.9605 | 0.9736 | 0.9798 |
| Air Bubble | Double UNet | -0.9817 | -0.9841 | 0.4060 | 0.4545 | 0.9997 | 0.9996 | 0.9641 | 0.9683 | 0.9817 | 0.9839 |
| Air Bubble | ResUNet++ | -0.9858 | -0.9873 | 0.4062 | 0.4550 | 0.9999 | 0.9999 | 0.9721 | 0.9745 | 0.9858 | 0.9871 |

*Table 13: Final ROC based comparison of segmentation models.*

| Tissue Artifact Type | UNet Architecture | ROC Score |
|---|---|---|
| Tissue Fold | DoubleUNet | 99.24% |
| Tissue Fold | ResUNet++ | 98.56% |
| Air Bubble | DoubleUNet | 1.00 |
| Air Bubble | ResUNet++ | 1.00 |

Table 12 demonstrates the accuracy (IOU and dice-coefficient), loss, precision and recall comparison for both train and validation dataset.

From table 13 we can see that, for air-bubble dataset, both model gains 100% accuracy in terms of area under curve or ROC score while in the case of tissue fold dataset, DoubleUNet outperforms ResUNet++.

## Comparison Graph

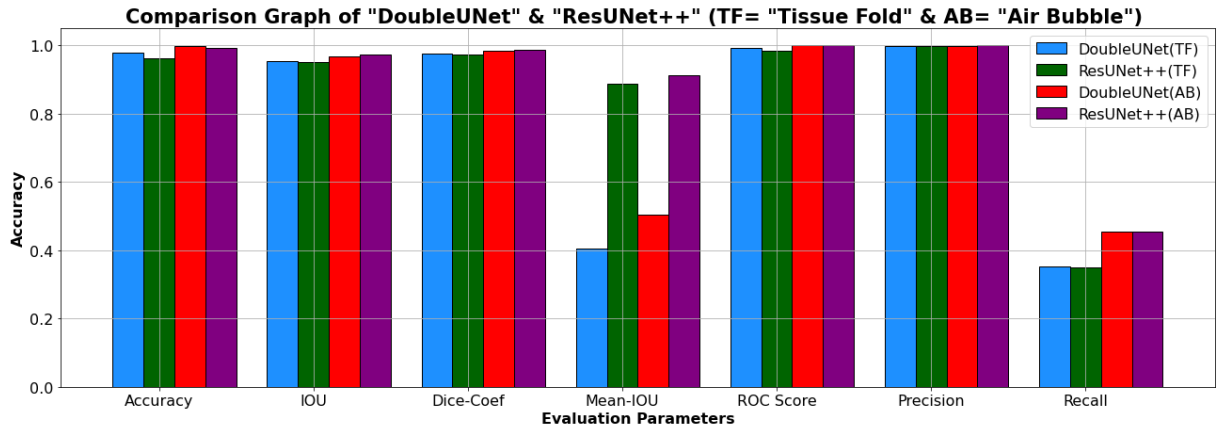

*Figure 17: Comparison graph of ResUNet++ and DoubleUNet.*

Over all The DoubleUNet out performs the ResUNet++. But the results are almost similar. Only the mean IOU value is greater in case of ResUNet++. But, Overall accuracy of DoubleUnet is better.

For the segmentation experiment, we can conclude that, our model works better for air-bubble dataset. Both models generate better result for air- bubble. While running 6-fold cross validation, in one or two cases, the result of ResUNet++ was better but overall the DoubleUNet works better which was explicitly clear throughout the experiment.

But one interesting findings was for both dataset resUNet++ generates better Mean-IOU than the DoubleUNet when the actual assessment criteria for our result was IOU and dice-coefficient.

Another finding from our segmentation experiment is that, the precision is over 99.9% while the recall is between 33%-45%. It means that, false positive rate is greater than false negative.

## 5.2. Severity Analysis/ Classification Result

Table 14 represents the accuracy and loss results of transfer learning based on 10 models.

*Table 14: Transfer Learning for base model selection.*

| Pretraind Model | Opimizer | Loss Function | Learning Rate | Validation Loss | Test Accuracy | ROC Score |
|---|---|---|---|---|---|---|
| Xception | Adam | CCE | 0.0001 | 0.0170 | 99.52% | 99.99% |
| | | KLD | 0.0001 | 0.0222 | 99.05% | 99.98% |
| | Adamax | CCE | 0.0001 | 0.0304 | 99.29% | 99.99% |
| | | KLD | 0.0001 | 0.0287 | 99.52% | 99.99% |
| | RMSprop | CCE | 0.0001 | 0.0103 | **99.76%** | **99.99%** |
| | | KLD | 0.0001 | 0.0093 | 99.52% | 100% |
| VGG16 | Adam | CCE | 0.0001 | 0.0444 | 98.81% | 99.99% |
| | | KLD | 0.0001 | 0.0480 | 98.57% | 99.98% |
| | Adamax | CCE | 0.0001 | 0.1173 | 97.86% | 99.93% |
| | | KLD | 0.0001 | 0.1192 | 97.86% | 99.92% |
| | RMSprop | CCE | 0.0001 | 0.0298 | 99.05% | 99.99% |
| | | KLD | 0.0001 | 0.0224 | 99.52% | 100% |
| VGG19 | Adam | CCE | 0.0001 | 0.0441 | 99.05% | 100% |
| | | KLD | 0.0001 | 0.0435 | 99.05% | 99.99% |
| | Adamax | CCE | 0.0001 | 0.1259 | 97.38% | 99.95% |
| | | KLD | 0.0001 | 0.1145 | 98.33% | 99.93% |
| | RMSprop | CCE | 0.0001 | 0.0217 | **99.76%** | **100%** |
| | | KLD | 0.0001 | 0.0252 | 99.05% | 99.99% |
| ResNet50 | Adam | CCE | 0.0001 | 0.4082 | 86.19% | 96.57% |
| | | KLD | 0.0001 | 0.4418 | 78.81% | 96.53% |
| | Adamax | CCE | 0.0001 | 0.5723 | 80.71% | 94.05% |
| | | KLD | 0.0001 | 0.5385 | 78.10% | 94.07% |
| | RMSprop | CCE | 0.0001 | 0.4903 | 78.57% | 95.77% |
| | | KLD | 0.0001 | 0.5587 | 80.71% | 95.12% |
| InceptionV3 | Adam | CCE | 0.0001 | 0.0422 | 99.05% | 99.96% |
| | | KLD | 0.0001 | 0.0457 | 99.05% | 99.96% |
| | Adamax | CCE | 0.0001 | 0.0492 | 99.05% | 99.89% |
| | | KLD | 0.0001 | 0.0523 | 99.05% | 99.84% |
| | RMSprop | CCE | 0.0001 | 0.0315 | 99.29% | 99.98% |
| | | KLD | 0.0001 | 0.0317 | 99.29% | 99.98% |
| Inception ResNetV2 | Adam | CCE | 0.0001 | 0.0299 | 98.81% | 99.99% |
| | | KLD | 0.0001 | 0.0269 | 99.05% | 99.98% |
| | Adamax | CCE | 0.0001 | 0.0631 | 98.81% | 99.92% |
| | | KLD | 0.0001 | 0.0441 | 98.81% | 99.95% |
| | RMSprop | CCE | 0.0001 | 0.0368 | 98.57% | 99.99% |
| | | KLD | 0.0001 | 0.0152 | 99.05% | 100% |
| MobileNet | Adam | CCE | 0.0001 | 0.0631 | 98.81% | 99.92% |
| | | KLD | 0.0001 | 0.0229 | 99.29% | 99.99% |
| | Adamax | CCE | 0.0001 | 0.0296 | 99.29% | 99.98% |
| | | KLD | 0.0001 | 0.0263 | 99.52% | 99.97% |
| | RMSprop | CCE | 0.0001 | 0.0174 | 99.76% | 99.99% |
| | | KLD | 0.0001 | 0.0093 | **99.76%** | **100%** |
| MobileNetV2 | Adam | CCE | 0.0001 | 0.0148 | 99.52% | 99.99% |
| | | KLD | 0.0001 | 0.0119 | **99.76%** | **99.99%** |
| | Adamax | CCE | 0.0001 | 0.0169 | 99.52% | 100% |
| | | KLD | 0.0001 | 0.0245 | 99.29% | 99.99% |
| | RMSprop | CCE | 0.0001 | 0.0024 | **99.76%** | **100%** |
| | | KLD | 0.0001 | 0.0207 | **99.76%** | **100%** |
| DenseNet121 | Adam | CCE | 0.0001 | 0.0296 | 98.81% | 99.99% |
| | | KLD | 0.0001 | 0.0212 | 99.29% | 99.99% |
| | Adamax | CCE | 0.0001 | 0.0280 | 99.52% | 99.99% |
| | | KLD | 0.0001 | 0.0406 | 98.81% | 99.96% |
| | RMSprop | CCE | 0.0001 | 0.0035 | **99.76%** | **100%** |
| | | KLD | 0.0001 | 0.0171 | 99.52% | 99.99% |
| NasNetLarge | Adam | CCE | 0.0001 | 0.1059 | 98.81% | 99.92% |
| | | KLD | 0.0001 | 0.0971 | 98.81% | 99.97% |
| | Adamax | CCE | 0.0001 | 0.0442 | 98.81% | 99.93% |
| | | KLD | 0.0001 | 0.0328 | 99.05% | 99.99% |
| | RMSprop | CCE | 0.0001 | 0.0799 | 99.05% | 99.93% |
| | | KLD | 0.0001 | 0.0387 | 99.29% | 99.99% |

This table represents the results of different combinations of transfer learning. This table is used for base model selection for ensemble learning. The base models are selected comparing the test accuracy. There are 7 models which have same test accuracy. Among those 7 models, 1 model has relatively higher validation loss. So, we excluded that model from our base model list. So, we finally selected the top 6 models as our base model for applying ensemble learning. The top models are described in details in the next page.

Every base model selected for ensemble learning is described in details in terms of their loss, accuracy, miss-classification rate and ROC score in their respective pages along with the confusion matrix.

## Best Models

**Model 1**

| Model | XCEPTION |
|---|---|
| Epoch | 25 |
| Batch size | 32 |
| Learning rate | 0.0001 |
| Optimizer | RMSprop |
| Loss function | Categorical cross entropy (CCE) |

| Loss | Test accuracy | ROC Score | Mismatch |
|---|---|---|---|
| 0.0103149469 | 0.9976190328 | 0.9999914965 | 1 of 420 (1 mid predicted as high) |

**Graph**

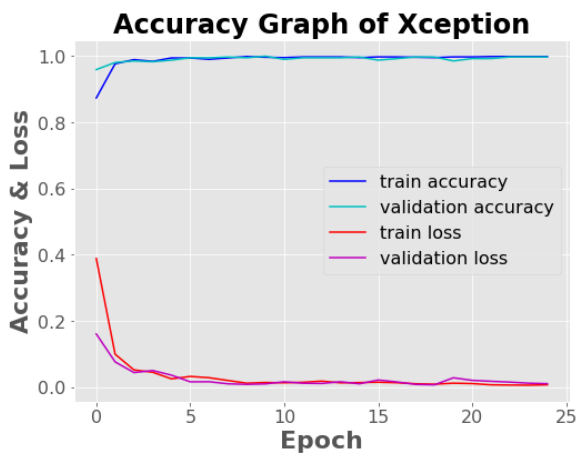
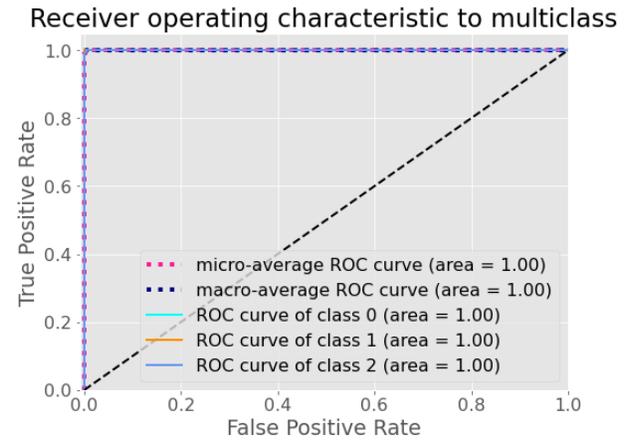

**Confusion Matrix**

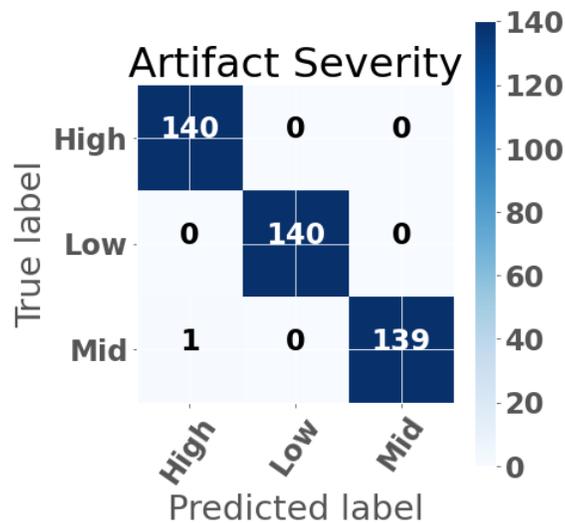

**Model 2**

| Model | VGG19 |
|---|---|
| Epoch | 25 |
| Batch size | 32 |
| Learning rate | 0.0001 |
| Optimizer | RMSprop |
| Loss function | Categorical cross entropy (CCE) |

| Loss | Test accuracy | ROC Score | Mismatch |
|---|---|---|---|
| 0.0216745790 | 0.9976190328 | 1.0000000000 | 1 of 420 (1 low predicted as high) |

**Graph**

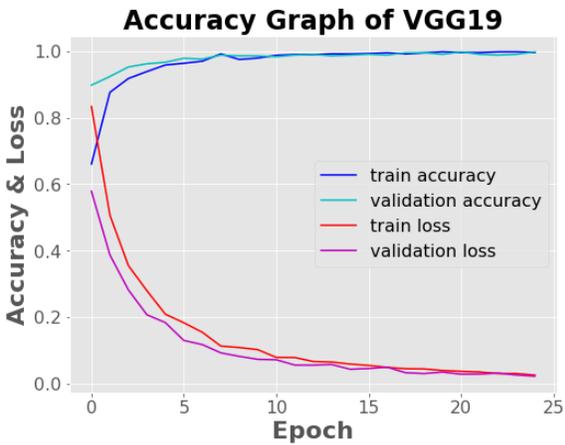
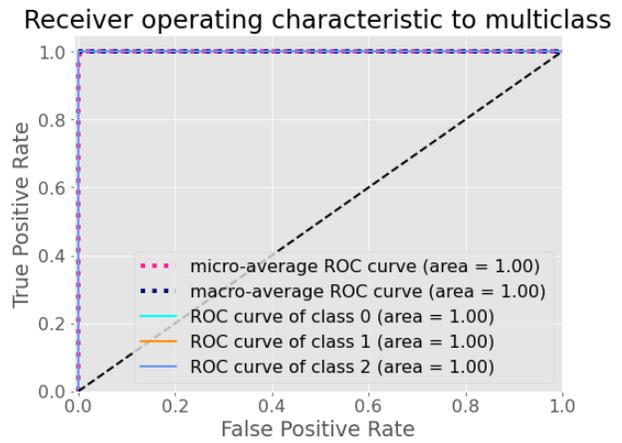

**Confusion Matrix**

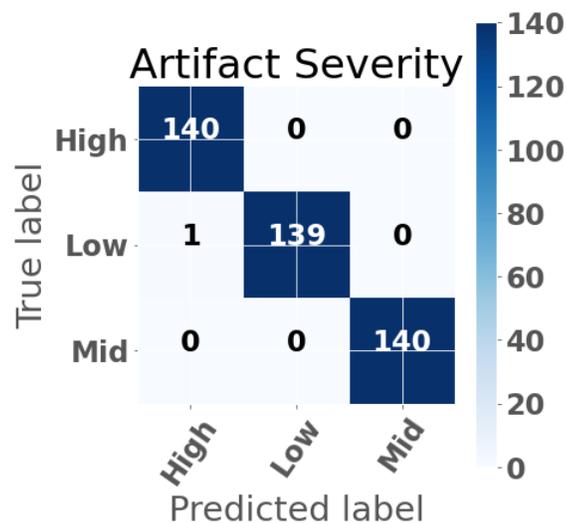

**Model 3**

| Model | MobileNet |
|---|---|
| Epoch | 25 |
| Batch size | 32 |
| Learning rate | 0.0001 |
| Optimizer | RMSprop |
| Loss function | Kullback Leibler Divergence (KLD) |

| Loss | Test accuracy | ROC Score | Mismatch |
|---|---|---|---|
| 0.0092747136 | 0.9976190328 | 1.0000000000 | 1 of 420 (1 mid predicted as high) |

**Graph**

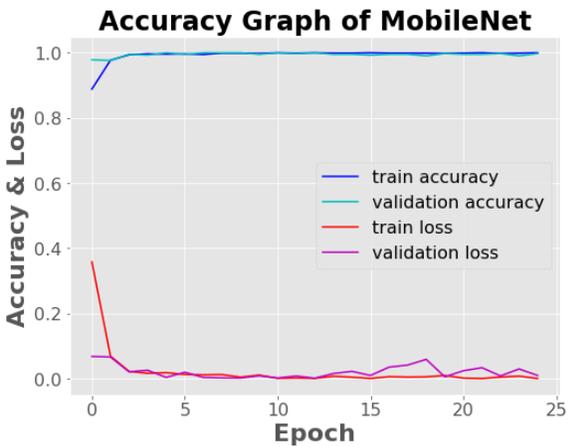 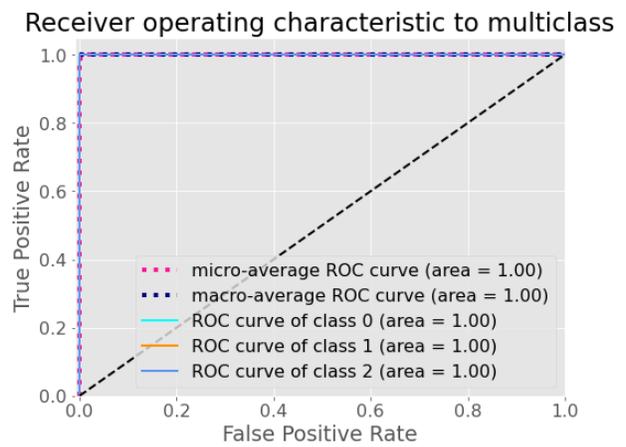

**Confusion Matrix**

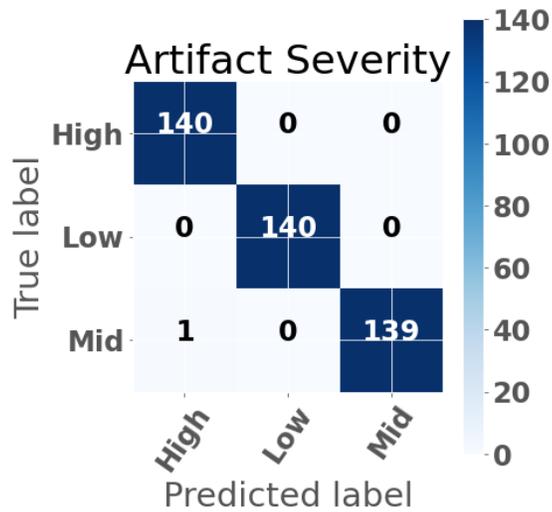

**Model 4**

| Model | MobileNetV2 |
|---|---|
| Epoch | 25 |
| Batch size | 32 |
| Learning rate | 0.0001 |
| Optimizer | RMSprop |
| Loss function | Categorical cross entropy (CCE) |

| Loss | Test accuracy | ROC Score | Mismatch |
|---|---|---|---|
| 0.0024069594 | 0.9976190328 | 1.0000000000 | 1 of 420 (1 mid predicted as high) |

**Graph**

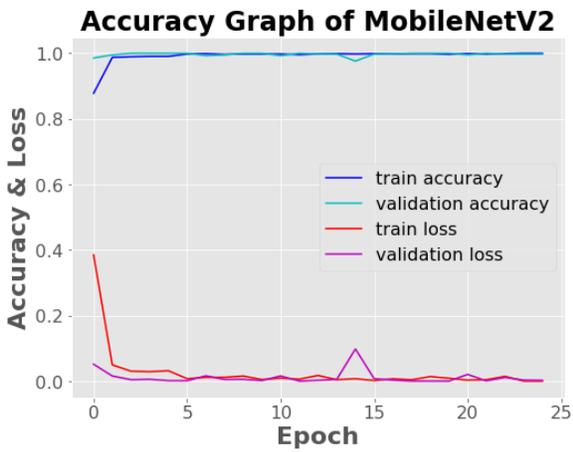
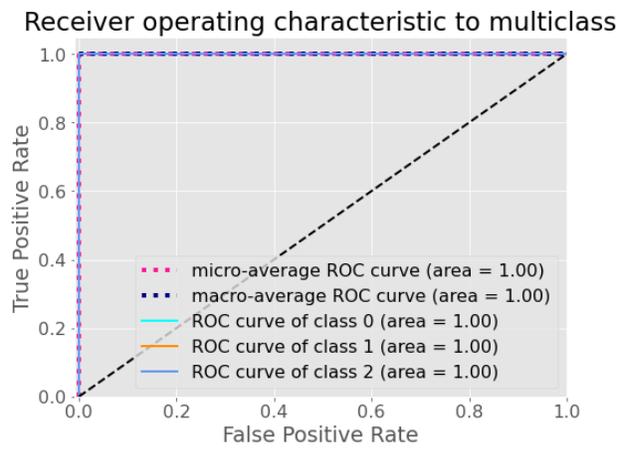

**Confusion Matrix**

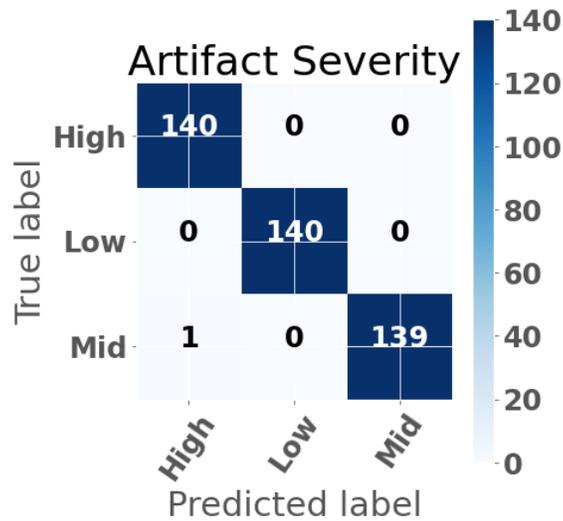

**Model 5**

| Model | DenseNet121 |
|---|---|
| Epoch | 25 |
| Batch size | 32 |
| Learning rate | 0.0001 |
| Optimizer | RMSprop |
| Loss function | Categorical cross entropy (CCE) |

| Loss | Test accuracy | ROC Score | Mismatch |
|---|---|---|---|
| 0.0034660110 | 0.9976190328 | 1.0000000000 | 1 of 420 (1 mid predicted as high) |

**Graph**

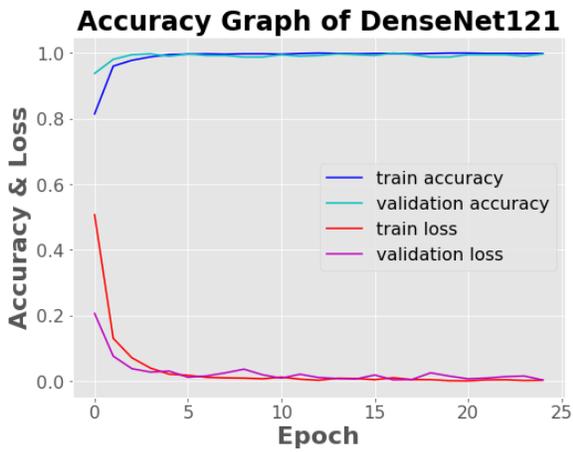
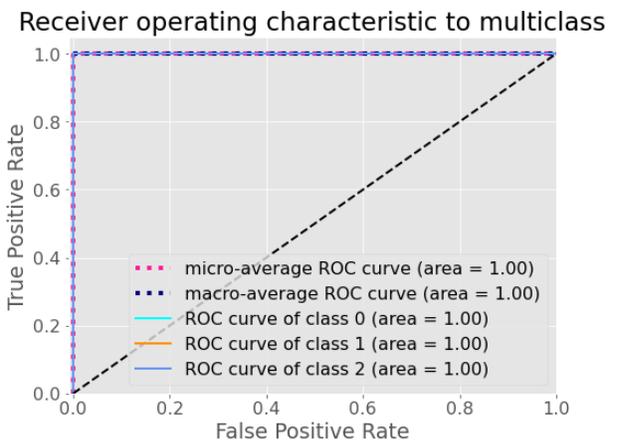

**Confusion Matrix**

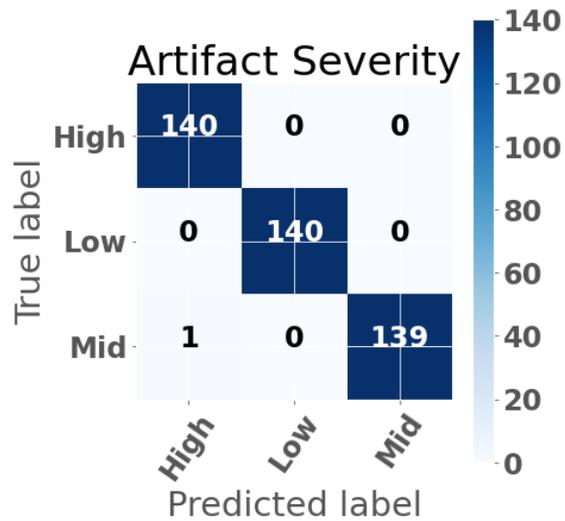

**Model 6**

| Model | MobileNetV2 |
|---|---|
| Epoch | 25 |
| Batch size | 32 |
| Learning rate | 0.0001 |
| Optimizer | RMSprop |
| Loss function | Kullback Leibler Divergence (KLD) |

| Loss | Test accuracy | ROC Score | Mismatch |
|---|---|---|---|
| 0.0206717420 | 0.9976190328 | 1.0000000000 | 1 of 420 (1 mid predicted as high) |

**Graph**

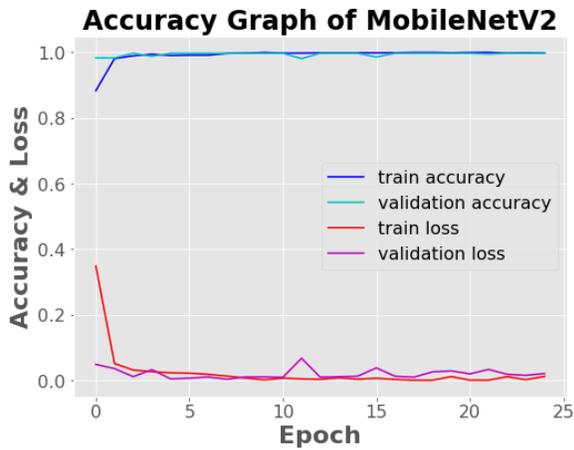
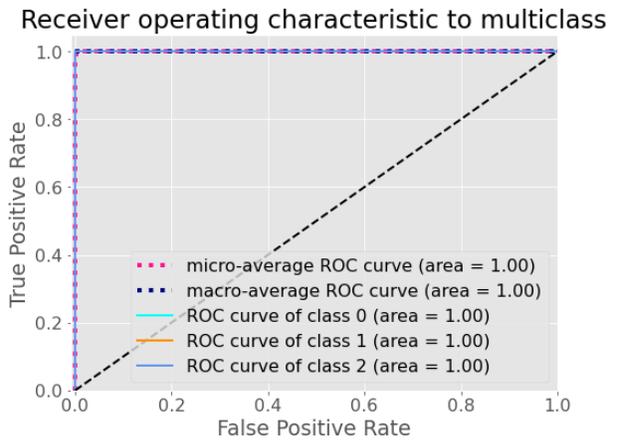

**Confusion Matrix**

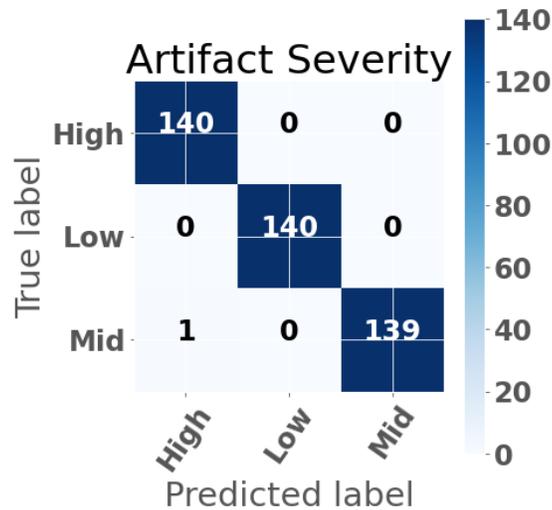

*Table 15: Top 6 models description*

| Model name | Combination | | | Loss | Accuracy |
|---|---|---|---|---|---|
| | **Pretrained Model** | **Optimizer** | **Loss function** | | |
| Model 1 | XCEPTION | RMSprop | CCE | 0.0103149469 | 0.9976190328 |
| Model 2 | VGG19 | RMSprop | CCE | 0.0216745790 | 0.9976190328 |
| Model 3 | MobileNet | RMSprop | KLD | 0.0092747136 | 0.9976190328 |
| Model 4 | MobileNetV2 | RMSprop | CCE | 0.0024069594 | 0.9976190328 |
| Model 5 | DenseNet121 | RMSprop | CCE | 0.0034660110 | 0.9976190328 |
| Model 6 | MobileNetV2 | RMSprop | KLD | 0.0206717420 | 0.9976190328 |

For Ensemble Learning various methods can be applied such as stacking, voting etc. For our experiment, we have used stacking.

We used various combinations of models as follows:

- **Using top 6 models**
- **Using top 5 models**
- **Using top 4 models**
- **Using top 3 models**
- **Using top 2 models**

And for every combination, we used Logistic Regression, KNN, SVM, Decision Tree, Random Forest, Ada Boost Classifier, XGB Classifier, GB Classifier, GB Regressor and Gaussian Naïve bias as meta learners and compared the results. The comparison table is mention in the next page.

Table 16: *Ensemble Learning Combinations using 10 meta learners.*

| Models | Combination | Accuracy (%) | | | | | | | | | |
|---|---|---|---|---|---|---|---|---|---|---|---|
| | | Logistic Regression | KNN | SVM | Decision Tree | Random Forest | AdaBoost Classifier | XGB Classifier | GB Regressor | GB Classifier | Gaussian NB |
| Top 2 | Model (4, 5) | 99.9999 | 99.9999 | 99.9999 | 99.8211 | 99.9999 | 99.9999 | 99.9999 | 99.9999 | 99.9999 | 99.9999 |
| Top 3 | Model (3, 4, 5) | 99.9999 | 99.9999 | 99.9999 | 99.9999 | 99.9999 | 99.9999 | 99.9999 | 99.9999 | 99.9999 | 99.9999 |
| Top 4 | Model (1, 3, 4, 5) | 99.9999 | 99.9999 | 99.9999 | 99.6441 | 99.8211 | 99.9999 | 99.8217 | 99.8217 | 99.9999 | 99.9999 |
| Top 5 | Model (1, 2, 3, 4, 5) | 99.9999 | 99.9999 | 99.9999 | 99.6415 | 99.9999 | 99.9999 | 99.8211 | 99.9999 | 99.8211 | 99.9999 |
| Top 6 | Model (1, 2, 3, 4, 5, 6) | 99.9999 | 99.9999 | 99.9999 | 99.8211 | 99.9999 | 99.9999 | 99.9999 | 99.8211 | 99.8217 | 99.8217 |

From the table it is clear that,

Logistic Regression, K Nearest Neighbor and Ada Boost Classifier performs better than any other models for our experiment. And from the model perspective, using the top 3 models generates the best result.

## 5.3. Critical Analysis

We have so far covered the results and related discussion on both the segmentation and classification tasks. In terms of the segmentation part, the segmentation accuracy is assessed based on some particular IOU threshold value instead of pixel wise accuracy. If the value is above 90% we considered it to be as a successful identification case. Based on this assessment criterion the segmentation result we got is quite good which is over 97% (while using DoubleUNet) for air bubble and over 96% (while using DoubleUNet) for tissue fold. Due to manual binary masking, some masking imperfection occurred as it couldn't have been avoided altogether. But, if perfect binary masking was possible, then we could get even better results in terms of segmentation accuracy.

As for classification task, the levels of severity were determined by human eye. For this, the result can very if the annotation is done by another human. But, looking at the result, which is 99.99%, we can easily say that, our model will definitely achieve 99% accuracy in terms of severity classification even if done by different experts providing that the binary masking should be close to perfect and the data annotation process should follow standard annotation techniques.

Another thing is that, for severity analysis, regression is better than classification as no hard and fast rules applies to the severity labels, but in that case, the manual observation will be required even after visualization of the artifacts and human decision will be necessary. In our case, to avoid that scenario we implemented classification instead of regression.

# 6.Conclusion

In this research, we worked on improving the tissue artifact segmentation accuracy to further enhance the detection mechanism of the whole slide imaging scanner. We have managed to achieve our goal using DoubleUNet and ResUNEt++ architecture; the DoubleUnet architecture performs better in most of the cases. We have also proposed an ensemble learning based classification model which analyze the severity of the detected artifact. Our model was tested on a dataset annotated by experts in this field. This will help the WSI to take decision on removing artifacts from diagnosis process according to the severity detected by our model. This proposed model will enhance the WSI experience and lead the digital pathology to a new level, making the diagnosis process easier than ever before and also more accurate.

Our future plan is to apply regression for severity analysis instead of classification and also to reconstruct the artifact containing region and to check whether the reconstructed area can be used in meaningful diagnosis or not. If these can be done successfully, the digital pathology will be elevated to a new step and the future of pathological image analysis will be brighter than ever before.

## Author Information

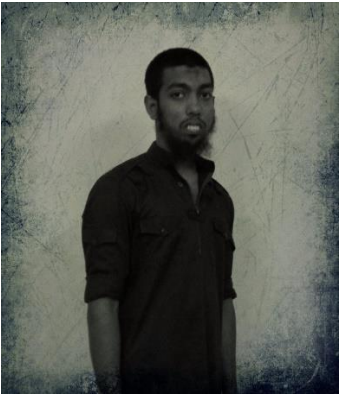

**Galib Muhammad Shahriar Himel**

**ID: 21-92094-2**

MSCS, Dept. of Computer Science, AIUB

ORCID: https://orcid.org/0000-0002-2257-6751

LinkedIn: https://www.linkedin.com/in/galib-muhammad-shahriar-himel-3763b0138/

Email: galib.muhammad.shahriar@gmail.com